\documentclass[11pt, a4paper]{article}

\usepackage{a4wide, amsmath, amssymb, amsthm}
\usepackage{tikz}
\usetikzlibrary{calc}

\tikzstyle{v} = [circle, draw, thick, inner sep=0pt, minimum size=2mm]
\tikzstyle{b} = [v, fill=blue]
\tikzstyle{r} = [v, fill=red]
\tikzstyle{s} = [v, fill=black] 
\tikzstyle{w} = [v, fill=white]
\tikzstyle{a} = [v, fill=yellow!50!brown] 

\usepackage{algorithmicx}
\usepackage{algorithm}
\usepackage[noend]{algpseudocode}
\usepackage[colorinlistoftodos]{todonotes}
\usepackage{soul}
\usepackage{url}
\definecolor{purple}{cmyk}{.51,.91,0,.34}
\usepackage{microtype}
\usepackage[pdftex, colorlinks,
  linkcolor=blue,
  citecolor=purple,
  filecolor=blue,urlcolor=blue]%
  {hyperref}

\newtheorem{theorem}{Theorem}

\usepackage{graphicx}
\usepackage{amsmath}
\usepackage{graphicx}
\usepackage{amssymb}
\usepackage{float}
\usepackage{todonotes}

\makeatletter
\renewcommand\subsubsection{\@startsection{subsubsection}{3}{\z@}%
                       {-18\p@ \@plus -4\p@ \@minus -4\p@}%
                       {0.5em \@plus 0.22em \@minus 0.1em}%
                       {\normalfont\normalsize\bfseries\boldmath}}
\makeatother
\usepackage{color}

\newtheorem{open-problem}[theorem]{Open Problem}

\newtheorem{lemma}[theorem]{Lemma}

\newtheorem{cor}[theorem]{Corollary}
\newtheorem{dfn}[theorem]{Definition}
\newtheorem{definition}[theorem]{Definition}

\newcommand{\CH}{C{\kern-1.5pt}H}
\newcommand{\DT}{\mathcal{D}\kern-.5pt\mathcal{T}}

\title{Interval $H$-graphs\,:\linebreak[1] Recognition and forbidden obstructions}

\author{
Haiko M\"{u}ller\thanks{School of Computer Science, University of Leeds, Leeds, UK. Email: \texttt{h.muller@leeds.ac.uk}}
\and 
Arash Rafiey\thanks{
Indiana State University, Indiana, USA.
Email: \texttt{arash.rafiey@indstate.edu} {, supported by Bailey Faculty Fellowship}
}}
\date{}

\begin{document}

\maketitle

\begin{abstract}
We introduce the class of \emph {interval $H$-graphs}, which is the generalization of interval graphs, particularly interval bigraphs. For a fixed graph $H$ with vertices $a_1,a_2,\dots,a_k$, we say 
that an input graph $G$ with given partition $V_1,\dots,V_k$ of its vertices is an interval $H$-graph
if each vertex $v \in G$ can be represented by an interval $I_v$ from a real line so that $u \in V_i$ 
and $v \in V_j$ are adjacent if and only if $a_ia_j$ is an edge of $H$ and  intervals $I_u$ and $I_v$ intersect.
$G$ is called interval $k$-graph if $H$ is a complete graph on $k$ vertices. 
and interval bigraph when $k=2$.  
We study the ordering characterization and forbidden obstructions of interval 
$k$-graphs and present a polynomial-time recognition algorithm for them. 
Additionally, we discuss how this algorithm can be extended to recognize general interval $H$-graphs. 
Special cases of interval $k$-graphs, particularly comparability interval $k$-graphs, 
were previously studied in \cite{brown}, where the complexity interval $k$-graph recognition was posed as an open problem.

\end{abstract}

\section{Introduction and Problem Definition}

The vertex set of a graph $G$ is denoted by $V(G)$ and the edge
set of $G$ is denoted by $E(G)$. A graph $G$ is called an \emph{interval graph}, if there exists a family $I_v$, $v \in V(G)$, of
intervals (from the real line) such that, for all different $x,y \in V(G)$
the vertices $x$ and $y$ are adjacent in $G$ if and
only if $I_x$ and $I_y$ intersect.  
A bigraph $G$ is a bipartite graph with a fixed bipartition into
\emph{black} and \emph{white} vertices. We sometimes denote these
sets as $B$ and $W$, and view the vertex set of $G$ as partitioned
into $(B,W)$. A bigraph
$G$ is called an \emph{interval bigraph} if there exists a family $I_v$, $v \in B \cup W$, of
intervals (from the real line) such that, for all $x \in B$
and $y \in W$, the vertices $x$ and $y$ are adjacent in $G$ if and
only if $I_x$ and $I_y$ intersect. Then, this family of intervals is
called an \emph{interval representation} of the bigraph $G$.

Interval bigraphs were introduced in \cite{hkm} and have been
studied in \cite{denver,hh2003,muller}. They are closely related
to interval digraphs introduced by Sen \textsl{et al.}~\cite{sdrw}. 
Interval bigraphs and interval digraphs have become of interest in such new areas
 as graph homomorphisms, \textsl{e.g.}~\cite{adjust}.
 
A \emph{co-circular arc bigraph} is a bipartite graph whose complement
is a circular arc graph (see \cite{bls} for the definitions of graph classes not introduced here).
The class of interval bigraphs is a subclass of co-circular arc bigraphs. Indeed, the former class consists exactly of those bigraphs whose complement is the intersection of a family of circular arcs no two of which
cover the circle \cite{hh2003}. There is a linear-time recognition algorithm for
co-circular arc bigraphs \cite{ross}. On the other hand, the class of interval bigraphs is a
super-class of proper interval bigraphs (also known as bipartite permutation graphs \cite{bss} or monotone graphs \cite{DyMu19}), 
for which there is also a linear-time a linear time recognition algorithm \cite{hh2003,bss}.

Interval bigraphs can be recognized in polynomial time using the algorithm developed by M\"{u}ller \cite{muller}.  M\"{u}ller's
algorithm runs in time $\mathcal{O}(nm^6(n+m)\log n)$ where $m$ is the number of edges and $n$ is the number of vertices on input bigraph $G$. A faster algorithm was developed in \cite{arash-jgt}; with running time $\mathcal{O}(mn)$. But there are several linear time algorithms for recognition of \emph{interval  graphs}, are known, \textsl{e.g.}, \cite{booth,corneil,corneil09,habib,korte}.

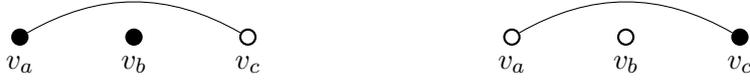
\begin{figure}[hbtp]
  \hspace*{\fill}
  \begin{tikzpicture}[scale=1.5]
    \node[s, label=below:$v_a$] (a) at (1,1) {}; 
    \node[s, label=below:$v_b$] (b) at (2,1) {}; 
    \node[w, label=below:$v_c$] (c) at (3,1) {};
    \draw (a) to[bend left] (c);
  \end{tikzpicture}
  \hspace*{\fill}
  \begin{tikzpicture}[scale=1.5]
    \node[w, label=below:$v_a$] (a) at (1,1) {}; 
    \node[w, label=below:$v_b$] (b) at (2,1) {}; 
    \node[s, label=below:$v_c$] (c) at (3,1) {};
    \draw (a) to[bend left] (c);
  \end{tikzpicture}
  \hspace*{\fill}
  \caption{Forbidden patterns for interval bigraphs}
  \label{fig:forbidden-pattern}
\end{figure}

We  use the ordering characterization of interval bigraphs in \cite{hh2003}. A bigraph $G$ is an interval bigraph
if and only if its vertices admit a linear ordering $<$
without any of the forbidden patterns in Figure~\ref{fig:forbidden-pattern}. Hence, we will rely on the existence of a linear ordering $<$ such that if $v_a < v_b <
v_c$ (not necessarily consecutively) and $v_a,v_b$ have the same color and opposite to the color
of $v_c$ then $v_av_c \in E(G)$ implies that $v_bv_c \in E(G)$.

We call such an ordering $<$ a \emph{bi-interval ordering} for $G$.
%
%
%
There are several
graph classes that can be characterized by the existence of an ordering
without a number of forbidden patterns. One such class is the class of interval graphs. A graph $G$ is an interval graph if and only if
there exists an ordering $<$ of $V(G)$ such that none
of the following patterns appears \cite{dama,dgr}.

\begin{itemize}
\item $v_a < v_b < v_c$, $v_av_c, v_bv_c \in E(G)$ and $v_av_b \notin E(G)$
\item $v_a < v_b < v_c$, $v_av_c \in E(G)$ and $v_bv_c, v_av_b \notin E(G)$
\end{itemize}


Some of the other classes of graphs that have ordering characterizations without forbidden patterns are proper 
interval graphs, comparability graphs, co-comparability graphs, chordal graphs, convex bipartite graphs, co-circular arc bigraphs, permutation bigraphs, and
proper interval bigraphs \cite{esa-2014}. We, in particular, mention the ordering characterization of permutation bigraphs and co-circular arc bigraphs without forbidden patterns.

\begin{figure}[htbp] 
  \hspace*{\fill}
  \begin{tikzpicture}[scale=1.0, label distance=-3pt]
    \node[s, label=below:\strut$a_i$]    (a) at (1,1) {}; 
    \node[s, label=below:\strut$a_{i'}$] (b) at (2,1) {}; 
    \node[w, label=below:\strut$b_{j'}$] (c) at (3,1) {};
    \node[w, label=below:\strut$b_j$]    (d) at (4,1) {};
    \draw (a) to[bend left] (d);
    \draw (b) to[bend left] (c);
  \end{tikzpicture}
  \hspace*{\fill}
  \begin{tikzpicture}[scale=1.0, label distance=-3pt]
    \node[s, label=below:\strut$a_i$]    (a) at (1,1) {}; 
    \node[s, label=below:\strut$a_{i'}$] (b) at (2,1) {}; 
    \node[w, label=below:\strut$b_{j'}$] (c) at (3,1) {};
    \node[w, label=below:\strut$b_j$]    (d) at (4,1) {};
    \draw (a) to[bend left] (d);
    \draw (b) to[bend left] (d);
    \draw (b) to[bend left] (c);
  \end{tikzpicture}
  \hspace*{\fill}
   \caption{Forbidden Patterns for co-circular arc bigraphs}
   \label{fig:min-patterns}
 \end{figure}
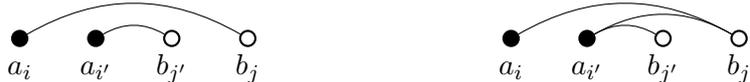

A bigraph $G=(A,B,E)$ is a co-circular arc bigraph if there is a linear ordering $a_1<\dots<a_p< b_1<b_2 < \dots <b_q$ (with $A=\{a_1,\dots,a_p\}, B=\{b_1,\dots,b_q\}$) so that if $a_ib_j,a_{i'}b_{j'}\in E$ with $i<i'$ and $j'<j$ then $a_ib_{j'} \in E$. Such an ordering $<$ is called a \emph{min ordering} of $G$ \cite{esa-2012}. The forbidden pattern corresponding to a min ordering are given in Figure~\ref{fig:min-patterns}.

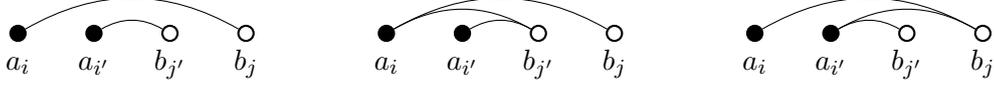
\begin{figure}[htbp] 
  \hspace*{\fill}
  \begin{tikzpicture}[scale=1.0, label distance=-3pt]
    \node[s, label=below:\strut$a_i$]    (a) at (1,1) {}; 
    \node[s, label=below:\strut$a_{i'}$] (b) at (2,1) {}; 
    \node[w, label=below:\strut$b_{j'}$] (c) at (3,1) {};
    \node[w, label=below:\strut$b_j$]    (d) at (4,1) {};
    \draw (a) to[bend left] (d);
    \draw (b) to[bend left] (c);
  \end{tikzpicture}
  \hspace*{\fill}
  \begin{tikzpicture}[scale=1.0, label distance=-3pt]
    \node[s, label=below:\strut$a_i$]    (a) at (1,1) {}; 
    \node[s, label=below:\strut$a_{i'}$] (b) at (2,1) {}; 
    \node[w, label=below:\strut$b_{j'}$] (c) at (3,1) {};
    \node[w, label=below:\strut$b_j$]    (d) at (4,1) {};
    \draw (a) to[bend left] (c);
    \draw (a) to[bend left] (d);
    \draw (b) to[bend left] (c);
  \end{tikzpicture}
  \hspace*{\fill}
  \begin{tikzpicture}[scale=1.0, label distance=-3pt]
    \node[s, label=below:\strut$a_i$]    (a) at (1,1) {}; 
    \node[s, label=below:\strut$a_{i'}$] (b) at (2,1) {}; 
    \node[w, label=below:\strut$b_{j'}$] (c) at (3,1) {};
    \node[w, label=below:\strut$b_j$]    (d) at (4,1) {};
    \draw (a) to[bend left] (d);
    \draw (b) to[bend left] (d);
    \draw (b) to[bend left] (c);
  \end{tikzpicture}
  \hspace*{\fill}
   \caption{Forbidden Patterns for bipartite permutation graphs (proper interval bigraphs)}
   \label{fig:min-max-patterns}
 \end{figure}

Likewise, a bigraph $G=(A,B,E)$ is a permutation bigraph if there is a linear ordering $a_1<\dots<a_p < b_1<b_2 < \dots <b_q$ (with $A=\{a_1,\dots,a_p\}, B=\{b_1,\dots,b_q\}$) so that if $a_ib_j,a_{i'}b_{j'}\in E$ with $i<i'$ and $j'<j$ then $a_ib'_j,a'_ib_j\in E$. Such an ordering $<$ is called a \emph{min-max ordering} of $G$ \cite{gutin,monotone}. The forbidden pattern corresponding to a min-max ordering are given in Figure~\ref{fig:min-max-patterns}. 

Let $G=(A,B,E)$ be an interval bigraph. Let $x_1\prec \dots\prec x_n$ be a linear ordering of vertices of $G$ without the forbidden patterns in Figure~\ref{fig:forbidden-pattern}. Let $<$ be the reverse of $\prec$; that is, let $x_n<\dots<x_1$. Let $a_1 < a_2< \dots < a_p$ be the vertices in $A$ under ordering $<$, and $b_1<b_2<\dots<b_q$ be the vertices in $B$ under ordering $<$. Then, it is easy to see that $a_1<\dots<a_p< b_1<b_2 < \dots <b_q$
provides a min ordering for $G$. Thus, interval bigraphs are a subclass of co-circular arc bigraphs.

Similarly, let $G=(A,B,E)$ be a permutation bigraph, and let 
$a_1<\dots<a_p < b_1<\dots <b_q$ be a min-max ordering of $G$ (with $A=\{a_1,\dots,a_p\}$, $B=\{b_1,b_2,\dots,b_q\}$). Then, $<$  is a bi-interval ordering for $G$. Thus, permutation bigraphs are a subclass of interval bigraphs.

Following this line of research, we introduce a class of interval $ k$-graphs where we still have a total ordering of the vertices of the input graph $G$, but there are partitions of the vertices, and the forbidden patterns are defined with colored vertices. We consider the graph $G$, which admits a $k$-coloring with the given coloring (given partitions), and we say $G$ is an interval $k$-graph according to the following definition.

\begin{definition}[interval $k$-graphs]
Let $G$ be a $k$-partite graph ($k \ge 2$) with the given partite sets $V_1,V_2,\dots,V_k$. We say that $G$ is an \emph{interval $k$-graph} if there is a family of real line intervals $I_v$, $v \in V(G)$, so that for all $u,v \in V(G)$ from different partite sets, $uv \in E(G)$ if and only if $I_u,I_v$ intersect. 
\end{definition}

Herein, we identify $V_1,\dots,V_k$ with a $k$-coloring of $G$ and simply say that two vertices have different colors whenever they belong to different partite sets.

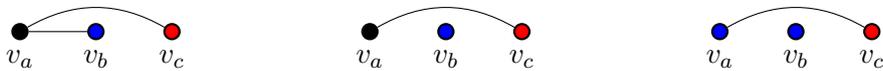
\begin{figure}[htbp]
  \hspace*{\fill}
  \begin{tikzpicture}[scale=1.0]
    \node[s, label=below:$v_a$] (a) at (1,1) {}; 
    \node[b, label=below:$v_b$] (b) at (2,1) {}; 
    \node[r, label=below:$v_c$] (c) at (3,1) {};
    \draw (a)--(b);
    \draw (a) to[bend left] (c);
  \end{tikzpicture}
  \hspace*{\fill}
  \begin{tikzpicture}[scale=1.0]
    \node[s, label=below:$v_a$] (a) at (1,1) {}; 
    \node[b, label=below:$v_b$] (b) at (2,1) {}; 
    \node[r, label=below:$v_c$] (c) at (3,1) {};
    \draw (a) to[bend left] (c);
  \end{tikzpicture}
  \hspace*{\fill}
  \begin{tikzpicture}[scale=1.0]
    \node[b, label=below:$v_a$] (a) at (1,1) {}; 
    \node[b, label=below:$v_b$] (b) at (2,1) {}; 
    \node[r, label=below:$v_c$] (c) at (3,1) {};
    \draw (a) to[bend left] (c);
  \end{tikzpicture}
  \hspace*{\fill}
  \caption{Forbidden patterns for interval $k$-graphs.}
  \label{fig:mix-patterns}
 \end{figure}

Notice that an interval graph $G$ is an interval $k$-graph where $k=|G|$. Thus, interval $k$-graphs generalize interval graphs. Let $G$ be a $k$-partite graph with the given partite sets $V_1,V_2,\dots,V_k$. We will show that $G$ is an interval $k$-graph if and only if $G$ admits an ordering $u_1<u_2<\dots< u_n$ of its vertices without the forbidden patterns depicted in Figure~\ref{fig:mix-patterns}. 

\begin{figure}[htbp]
  \begin{center}
    \begin{tikzpicture}[yscale=1.5, label distance=-3pt]
      \foreach \x in {1,2,3} {
        \pgfmathtruncatemacro{\z}{\x+3}
	      \node[s, label=below:\strut$\z$] (\z) at (\x,1) {};
	    \node[w, label=above:\strut$\x$] (\x) at (\x,2) {};
	    \draw (\x)--(\z);
      }
      \foreach[count=\x from 5] \u/\l in {a/d, b/e, c/f} {
	    \node[s, label=below:\strut$\l$] (\l) at (\x,1) {};
	      \node[w, label=above:\strut$\u$] (\u) at (\x,2) {};
	      \draw (\l)--(\u);
      }
      \foreach \u in {1,2,3} \foreach \l in {d,e,f} \draw (\l)--(\u); 
    \end{tikzpicture}
 \caption{\small{Exobicliques: 
 Here, $B=\{4,5,6,d,e,f\}$, $W=\{1,2,3,a,b,c\}$ and $B_1=\{d,e,f\}$, $W_1=\{1,2,3\}$ 
 and $B \setminus B_1=\{4,5,6\}$, $W \setminus W_1=\{a,b,c\}$}. }
 \label{fig:fig2}
   \end{center}
 \end{figure}
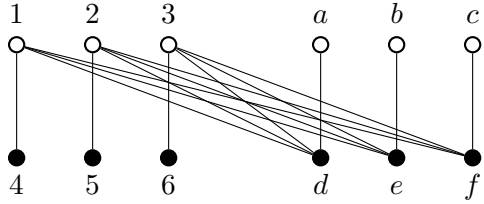

\paragraph{Obstruction for interval $k$-graphs:}
We have seen some of the forbidden obstructions of interval bigraphs in \cite{hh2003}. They are called \emph{exobicliques}. The bigraph $G=(B, W)$ is an exobiclique if the following hold. 
\begin{itemize}
 \item $B$ contains a nonempty part $B_1$ and $W$ contains a nonempty part 
$W_1$ such that $B_1 \cup W_1$ induces a biclique in $G$;
\item $B \setminus B_1$ contains three vertices with
incomparable neighborhood in $W_1$ and $W \setminus W_1$ contains three
vertices with incomparable neighborhoods in $B_1$ (an examples given in Figure~\ref{fig:fig2}).
\end{itemize}

\begin{figure}[htbp] 
 \begin{center}
    \begin{tikzpicture}[yscale=2.0, label distance=-2pt]
      \node[s, label=above:$v$] (v) at (6,3) {};
      \foreach[count=\x from 1] \u/\l in {x_0/y_0, y_1/x_1, w/w'} {
        \node[s, label=below:\strut$\l$] (\l) at (\x,1) {}; 
        \node[w, label=left:$\u$]        (\u) at (\x,2) {};
	    \draw (\l)--(\u)--(v);
      }
      \foreach[count=\x from 5] \u/\l in {y_2/x_2, x_3/y_3, u_0/v_0} {
        \node[s, label=below:\strut$\l$] (\l) at (\x,1) {}; 
        \node[w, label=above:$\u$]       (\u) at (\x,2) {};
	    \draw (\l)--(\u);
      }
      \foreach[count=\x from 9] \u/\l in {v_1/u_1, v_2/u_2, z/z'} {
        \node[s, label=below:\strut$\l$] (\l) at (\x,1) {}; 
        \node[w, label=right:$\u$]       (\u) at (\x,2) {};
	      \draw (\l)--(\u)--(v);
      }
      \foreach \l in {x_2, y_3, v_0} {
        \foreach \u in {x_0, y_1, w} \draw (\l)--(\u);
        \foreach \u in {v_1, v_2, z} \draw (\l)--(\u);
      }
    \end{tikzpicture}
 \caption{Forbidden Patterns}
 \label{fig:Obst3New}
 \end{center}
 \end{figure}
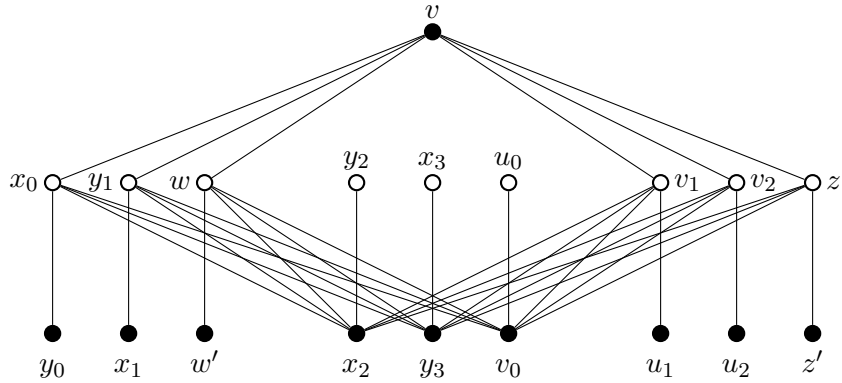

However, the obstruction for interval bigraphs (interval $2$-graphs) are not limited to exobiclique, there is a family of these obstruction considered in \cite{arash-jgt}. Figure \ref{fig:Obst3New} is an obstruction of the interval bigraph that is not an exobiclique.

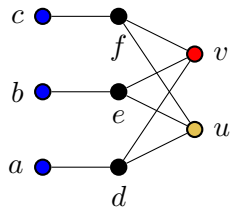
\begin{figure}[hbtp]
  \centering
  \begin{tikzpicture}[scale=1.0]
    \node[a, label=right:$u$] (u) at (3,1.5) {}; 
    \node[r, label=right:$v$] (v) at (3,2.5) {};
    \foreach[count=\y] \m/\n in {a/d, b/e, c/f} {
      \node[b, label=left:$\m$] (\m) at (1,\y) {}; 
      \node[s, label=below:$\n$] (\n) at (2,\y) {}; 
      \draw (\m)--(\n)--(u)  (\n)--(v);
    }
 \end{tikzpicture}
  \caption{Obstruction for interval $4$-graphs}
  \label{fig:so1}
\end{figure}

There are some new obstructions for interval $k$-graphs $k>2$, depicted in Figure \ref{fig:so1}.

\subsection{Our Results and Future Work}

Our primary contribution is the development of a recognition algorithm for interval $k$-graphs.

\begin{theorem}\label{th:main}
Let $G$ be a graph with a given partition of its vertices into $k$ partite sets $V_0, V_1, \dots,\linebreak[1] V_{k-1}$. Then, it can be determined in $O(|V(G)||E(G)|)$ time whether $G$ is an interval $k$-graph.
\end{theorem}

An interesting challenge arises when the $k$-coloring of $G$ is not provided, yet we still seek to determine whether $G$ is an interval $k$-graph. However, we do not have a conclusive indication that this problem becomes NP-complete when the $k$-coloring (a $k$-partition of its vertices) is not given. The authors of \cite{brown} investigated this problem for specific cases where $k = 2,3$. This leads us to propose the following open problem:

\begin{open-problem}\label{without-k-partition}
Can one determine in polynomial time whether a given graph $G$ admits a $k$-partition $V_1, V_2, \dots, V_k$ such that $G$, along with this partitioning, forms an interval $k$-graph?
\end{open-problem}

Identifying forbidden obstructions for interval $k$-graphs remains a significant challenge, even for $k=2$. As discussed in the previous subsection, the forbidden obstructions for interval bigraphs cannot be categorized into a finite number of families. 
This raises an important question for interval $k$-graphs:
However, it may be possible to enumerate them systematically. 

\begin{open-problem}
What are the forbidden obstructions for interval $k$-graphs?
\end{open-problem}

\section{Basic definitions and some preliminary results} 

Note that for $k>1$, a $k$-partite graph $G$ is an interval $k$-graph if and only if each
connected component of it is an interval $k$-graph. In the remainder of
this paper, we assume that $G = (V,E)$ is a connected $k$-partite graph with a
fixed partition $V_1,V_2,\dots,V_k$. By set of edges in the complete $k$-partite graph with
partite sets $V_1,\dots,V_k$ that are not present in $G$ we denote by
\[ \bar{E} = \{uv \mid u \in V_i, v \in V_j, 1 \le i < j \le k\} \setminus E\,. \]

We define \emph{pair-digraph} $G^+$ of $G$ corresponding to the forbidden patterns in Figure~\ref{fig:mix-patterns}, as follows. 
The set of vertexes of $G^+$ consists of all pairs $(u,v)$ such that $u, v \in V(G)$ and $u \neq v$. For clarity, we will often refer to vertices of $G^+$ as \emph{pairs} (in $G^+$). The arcs in $G^+$ are of one of the following types:

\begin{itemize}
 \item  $(u,v)(u',v)$ is an arc of $G^+$ when $u$ and $v$ belong to the same $V_i$ and $uu' \in E(G)$, and $vu' \notin E(G)$.

\item $(u,v)(u',v)$ is an arc of $G^+$ if $uu' \in E(G)$ and $u'v \in \bar{E}(G)$ and $u,v,v'$ all belong to different $V_i$.  

 \item $(u,v)(u,v')$ is an arc of $G^+$ when $u$ and $v'$ belong to the same $V_i$ with $vv' \in E(G) $, and $uv \notin E(G)$.

\item $(u,v)(u,v')$ is an arc of $G^+$ if $vv' \in E(G)$ and $uv \in \bar{E}(G)$ and $u,v,v'$ all belong to different $V_i$.  
\end{itemize}

Observe that if there is an arc from $(u,v)$ to $(u',v')$, then both $uv$ and $u'v'$ are non-edges of $G$. 
For two pairs $(x,y),(x',y') \in V(G^+)$ we say $(x,y)$ \emph{dominates} $(x',y')$ (or $(x',y')$ is \emph{dominated by} $(x,y)$) and  write $(x,y) \rightarrow (x',y')$ if there exists an arc (directed edge) from $(x,y)$ to $(x',y')$ in $G^+$. 
One should note that if $(x,y) \rightarrow (x',y')$ in $G^+$
then $(y',x') \rightarrow (y,x)$, to which property we will refer to as \emph{skew-symmetry}.\\

\noindent\textbf{Remark.} This idea of a pair-digraph can also be applied to the forbidden patterns in Figure  \ref{fig:min-patterns}. With $M=(A,B,E)$ being a connected bigraph, we define the \emph{pair-digraph} $M^*$ of $M$ corresponding to the forbidden pattern in Figure~\ref{fig:min-patterns} as follows. We set $V(M^*)=\{(u,v) \mid u,v\in A\text{ or } u,v\in B \} $ and $A(M^*)=\{(u,v)(u',v') \mid uu',vv'\in E,\; uv'\notin E\}$. Notice that if $(u,v)(u',v')\in A(M^*)$ then $(u',v')(u,v'),(u,v')(u,v)\in A(M^+)$. Then, it is easy to see that all vertices of each strong component of $M^*$ belong to the same strong component of $M^+$.

\begin{lemma}\label{newpairsap}
Let $<$ be an ordering of $G$ without the forbidden patterns in Figure~\ref{fig:mix-patterns}, and let $(u,v) \rightarrow (u',v')$ with $u<v$. Then, $u'<v'$.
\end{lemma}

\begin{proof} 

According to the definition of $G^+$, we either have
\begin{description}
    \item[Case (1)] $u,v$ have the same color, $v=v'$, $uu' \in E(G)$, and $vu' \notin E(G)$; or
    \item[Case (2)] $u,v$ have different colors, $u=u'$, $vv' \in E(G)$, and $uv \notin E(G)$
\end{description}
In Case (1) (resp.\ Case (2)), if $v'<u'$, then vertices $v',u,v$ (resp.\ $u,v,u')$--- in that order--- would induce a forbidden pattern in $G$, a contradiction. Hence, in both cases we will have $u'<v'$, as desired.
\end{proof}

We shall generally refer to a strong component of $G^+$ simply as a \emph{component} of $G^+$. We shall also identify a component by its vertex (pair) set. A component in $G^+$ is called \emph{non-trivial} if it contains more than one pair. For any component $S$ of $G^+$, we define its \emph{couple component}, denoted $S'$, to be $S' = \{(u,v):\ (v,u) \in S\}.$

The skew-symmetry property of $G^+$ implies the following fact.
\begin{lemma} \label{couplesap}
If $S$ is a component of $G^+$ then so is $S'$.
\end{lemma}
In light of Lemma \ref{couplesap}, for each component $S$ of $G^+$, $S$ and $S'$ are couple components of each other and we shall collectively refer to them as \emph{coupled components}. It can be easily shown that coupled components $S$ and $S'$ are either disjoint or equal. In the latter case, we say component $S$ is \emph{self-coupled}.

\begin{definition}[circuit]
For $n \ge 1$, a sequence $(x_0,x_1),(x_1,x_2),\dots, (x_{n-1},x_n)$, $(x_n,x_0)$ of pairs in 
a set $D \subseteq V(G^+)$ is called a \emph{circuit} in $D$.

\end{definition}

\begin{lemma}\label{rainap}
If a component of $G^+$ contains a circuit then $G$ is not an interval $k$-graph. 
\end{lemma}

Should say with respect to the fixed $k$ partition. 

\begin{proof} Let $(x_0,x_1),(x_1,x_2),\dots, (x_{n-1},x_n),(x_n,x_0)$ be a circuit in a component $S$ of $G^+$. Since $S$ is strongly connected, for all non-negative integers  $i$ and $j$ there exists a directed walk $W_{i,j}$ in $G^+$ from $(x_i,x_{i+1})$ to $(x_j,x_{j+1})$, where indices are mod $n+1$. Now, for all $i,j\ge 0$, following the sequence of pairs on $W_{i,j}$ and using Lemma \ref{newpairsap}, we conclude that $x_j<x_{j+1}$ whenever $x_i<x_{i+1}$. Hence, we must either have $x_i<x_{i+1}$ for all $i$, or $x_i>x_{i+1}$ for all $i$. However, since $x_{n+1}=x_0$, either case implies $x_0\not= x_0$; a contradiction. 
\end{proof}

If $G^+$ contains a self-coupled component then $G$ is not an interval $k$-graph. This is because a self-coupled component of $G^+$  contains two such pairs as $(u,v)$ and $(v,u)$, which comprise
a circuit of length $2$ (corresponding to $n=1$ in the definition of a circuit). However,  we will show that these are not the only obstructions to interval $k$-graphs (an in particular interval bigraphs), that is, there are bigraphs $G$ which are not interval bigraphs, despite $G^+$ not having any self-coupled component. In contrast, the obstructions to co-circular arc bigraphs are precisely the  components of $G^*$ containing both pairs $(x,y),(y,x)$.

\begin{theorem} \cite{esa-2012}
The bigraph $G$ is a co-circular arc bigraph if and only if it admits a min ordering and if and only if $G^*$ does not contain pairs $(x,y)$ and $(y,x)$ belonging to the same strong component of $G^*$. 
\end{theorem}


A \emph{tournament} is a directed graph that can be obtained from a complete undirected graph by
orienting each edge in one of the two possible directions.
A tournament is called \emph{transitive} if it is acyclic; i.e., if it does not contain a directed cycle. 

\begin{lemma}\label{turnajap}
Suppose that $G^+$ contains no self-coupled components, and let $D$ be any
subset of $V(G^+)$ containing exactly one component from each pair of coupled components.
Then, $D$ is the set of arcs of a tournament on $V(G)$.
Moreover, such a $D$ can be chosen to be a transitive tournament if and only if $G$ is an interval $k$-graph for any $k \ge 2$. 
\end{lemma}

In what follows, by a \emph{component} we mean a non-trivial
(strong) component unless we specify otherwise. For simplicity, we shall use a set $S$ of pairs in $G^+$
to also denote the sub-digraph of $G^+$ induced by $S$, when no confusion arises. 

We shall say two edges $ab$ and $cd $ of $G$ are \emph{independent} if the subgraph
of $G$ induced by the vertices $a, b, c,$ and $d$ has just the two edges $ab$ and $cd$. We shall say two disjoint induced subgraph $G_1$ and $G_2$ of $G$ 
are \emph{independent} if there is no edge of $G$ with one endpoint in $G_1$ and another endpoint in $G_2$.

First, to describe the algorithm, we introduce some technical definitions.

\begin{definition}[reachability closure]
Let $R$ be a subset of the pairs of $G^+$. Let $N^+[R]$ denote the set of all pairs in $G^+$ that are
reachable (via a directed path in $G^+$) from a pair in $R$. (Notice that $N^+[R]$ contains $R$.)
We call $N^+[R]$ the \emph{reachability closure} of $R$. We say that a pair $(u,v)$ is 
\emph{implied} by $R$ if $(u,v) \in N^+[R] \setminus R$. 
If $R=N^+[R]$, we say that $R$ is \emph{closed} under reachability. 
\end{definition}

\begin{definition}[envelope]
Let $R$ be a set of pairs of $G^+$. The \emph{envelope} of $R$, denoted $N^*[R]$, 
is the smallest set of pairs that contains $R$ and is closed under both reachability and transitivity
(if $(u,v),(v,w) \in N^*[R]$ then $(u,w) \in N^*[R]$). 
\end{definition}





\begin{lemma} \label{oneap}
Let $S$, $S'$ be the coupled components in $G^+$, so that both $N^*[S]$ and $N^*[S']$ contain a circuit. Then $G$ is not an interval $k$-graph.
\end{lemma}
\begin{proof}
According to Lemma \ref{turnajap} the final set $D$ must be a total ordering with
transitivity property. Therefore, one of $S$ and $S'$ must be in $D$. 
To find a total ordering that avoids the patterns in Figure~\ref{fig:forbidden-pattern},  
one of the $N^*[S]$, $N^*[S']$ must be in $D$, which is impossible. 
\end{proof}

We show the an ordering characterization of interval $k$-graphs. 
\begin{theorem}
For a fixed $k \ge 2$, let $G$ be a $k$-partite  graph with the given partite sets $V_1, V_2,
\dots, V_k$. $G$ is an interval $k$-graph if and only if $G$ admits an ordering $u_1<u_2<\dots< u_n$ of its vertices without the forbidden patterns depicted in Figure~\ref{fig:mix-patterns}. 
\end{theorem}

\begin{proof}
We denote the right and left end-points of an interval $I$ by $r(I)$ and $\ell(I)$, respectively.

First, suppose there is an interval representation $I_v, \;v\in G$ of $G$. Now, consider the total ordering $v _1 < v_2 < \dots < v_n$ of $G$ where for all $i,j$ we have $v_i<v_j$ when either $r(I_{v_i}) < r( I_{v_j})$, or $r(I_{v_i}) = r( I_{v_j})$ and  $\ell(I_{v_i}) \le \ell(I_{v_j})$. (In other words, we have $v_i<v_j$ whenever $(r(I_{v_i}),\ell(I_{v_i}))< (r(I_{v_j}),\ell(I_{v_j}))$ in the lexicographic ordering of pairs of real numbers.)

Now consider three indexes $a< b<c$. Assume $v_av_c$ is an edge of $G$, and vertices $v_b$ and $v_c$ have different colors. Since $v_av_ c \in E(G)$, $I_{v_a}$ and $I_{v_c}$ intersect and, hence,  
$\ell(I_{v_c}) < r(I_{v_a}) $. Moreover, since $v_a < v_b$, we have $r(I_{v_a}) \le r(I_{v_b})$. Therefore,  $\ell(I_{v_c}) < r(I_{v_b})$; i.e., $I_{v_b}$ and $I_{v_c}$ intersect. Thus, $v_bv_c$ is an edge of $G$, implying that none of the forbidden patterns in Figure~\ref{fig:mix-patterns} occurs.

In contrast, let $v_1<v_2< \dots < v_n$ be an ordering of the vertices of $G$ without the forbidden patterns in Figure~\ref{fig:mix-patterns}. For each $i$ set $r(J_{v_i})=i$ and 
$\ell(J_{v_i})=\min(\{i\} \cup \{j:\; v_j<v_i,\;v_iv_j \in E(G)\})$. 
One can easily see that $J_v,\;v\in G$, is an interval representation for $G$. 
\end{proof}
 


One can observe that if $u$ and $v$ are two vertices of $G$ so that they have in-comparable neighborhoods then $(u,v)$ is in a component of $G^+$. Indeed, with $uu',vv' \in E(G)$  and $uv',u'v \notin E(G)$ we get $(u,v) \rightarrow (u',v) \rightarrow (u',v') \rightarrow (u,v') \rightarrow (u,v)$. Using the same reasoning, if a pair $(x,y)$ outside component $S$ is dominated by a vertex in $S$, then $N(x) \subseteq N(y)$. 






\section{Structural Properties of Strong components in \boldmath$G^+$}
The following Lemma follows from the definition of $G^+$. 
\begin{lemma}\label{lem:two-independent-edges}
If $uu'$ and $vv'$ are independent edges in $G$ then the pairs $(u,v)$, $(u',v)$, $(u',v')$, and $(u,v')$ form a directed four-cycle of $G^+$ in the given order (resp.\ in reverse order). In particular, $(u,v)$, $(u',v)$, $(u',v')$, and $(u,v')$ belong to the same component of $G^+$.

\end{lemma}


\begin{lemma}\label{lem:strong-component}
Let $S$ be a component of $G^+$ containing a pair $(u,v)$ then one of the following occurs.
\begin{enumerate}
    \item $uv \in E(G)$ or $u$ and $v$ have the same colors and there exist $u',v'$ where $uu', vv'$ are edges of $G$, $uv',vu' \notin E(G)$. Furthermore, the four pairs $(u,v),
(u,v'), $   $(u',v)$, and $(u',v')$ are contained in $S$.
\item $u$ and $v$ have different colors and $uv \notin E(G)$ and there exists $u'$ such that $uu' \in E(G)$, $vu' \notin E(G)$ and $u,u',v$ all have different colors. 
\end{enumerate}
\end{lemma}
\begin{proof}
Since $S$ is a component, $(u,v)$ dominates some pair of $S$ and is
dominated by some pair of $S$. Firstly, suppose that $u$ and $v$ have the same color in $G$. 
Then $(u,v)$ dominates some $(u',v) \in S$ and is dominated by some $(u,v') \in S$. 
Now $uu'$ and $vv'$ must be edges of $G$, and $uv, uv', u'v,$ and $u'v'$ must be non-edges of $G$. 
Thus, $uu'$ and $vv'$ are independent edges in $G$. 
In this case, according to Lemma \ref{lem:two-independent-edges}, $S$ contains the directed cycle
$(u,v) \rightarrow (u',v) \rightarrow (u',v') \rightarrow (u,v') \rightarrow (u,v)$. 

Secondly, suppose $uv \in E(G)$. Now there must be a pair $(u',v) \in S$ dominated by $(u,v)$ 
where in this case we have $uu' \in E(G)$, $uv' \notin E(G)$, and $u,v'$ have different colors. 
Analogously, there must be some pair $(u,v') \in S$ that dominates $(u,v)$ where in this case 
we have $vv' \in E(G)$, $u'v \notin E(G)$, and $u',v$ have different colors. 
Now $(u,v) \rightarrow (u',v) \rightarrow (u',v') \rightarrow (u,v') \rightarrow (u,v)$. 

Finally, suppose that $u$ and $v$ have different colors. We note that $(u,v)$ dominates some
$(u,v') \in S$, and hence, $uv \notin E(G)$ and $vv'$ is
an edge of $G$. Now, if $u,v,v'$ have different colors, then $(u,v')$ also dominates $(u,v)$, implying that $(u,v),(u,v')$ are in the same component $S$.
\end{proof}

The structure of the components of $G^+$ is quite special,
and the trivial components interact with them in simple ways. A trivial component will be called 
a \emph{source} if its unique pair has in-degree zero, and 
a \emph{sink} if its unique pair has out-degree zero. 
Herein, we further explore these properties by establishing several lemmas. 
To do this, we need the following definition of the reachability of pairs in $G^+$.  

\begin{dfn}[reachability closure]
Let $R$ be a subset of the pairs of $G^+$. Let $N^+[R]$ denote the set of all pairs in $G^+$ that are
reachable (via a directed path in $G^+$) from a pair in $R$. (Notice that $N^+[R]$ contains $R$.) 
We call $N^+[R]$ the \emph{reachability closure} of $R$. We say that a pair $(u,v)$ is \emph{implied} 
by $R$ if $(u,v) \in N^+[R] \setminus R$. If $R=N^+[R]$, we say that $R$ is \emph{closed} under reachability.   
\end{dfn}

\begin{lemma}\label{implyap}
A pair $(a,c)$ is in $N^+[S] \setminus S$ (implied by $S$) for some component $S$ of $G^+$ if one of the following occurs. 

\begin{enumerate}
\item $a$ and $c$ have the same color. $N(a) \subseteq N(c)$, and there exist $bd,dc \in E(G)$ 
so that $a,b,d$ all have different colors and $ab,ad \notin E(G)$, and $(a,b),(a,d)$ are in $S$.  
\item $a$ and $c$ have the same color. $N(a) \subseteq N(c)$ and $G$ contains path $a, b, c, d, e$,
such that  and $ad,be \notin E(G)$, and  $(a,d), (a,e), (b,d),(b,e)$ lie in $S$.
\item $ac \in E(G)$ and $G$ contains a path $b,a,c,d$ such that $N(a) \setminus \{c\} \subseteq N(c)$ and $ad,bd \notin E(G)$, $(a,d), (b,d)$ lie in $S$.
\item $ac \in E(G)$ and $G$ contains path $b,d,c$ such that $N(a) \setminus \{c\} \subseteq N(c)$, and $ab,ad \notin E(G)$ and $a,d$ have different colors and $a,b$ have different colors. Furthermore, $(a,d),(a,b) \in S$. 
\end{enumerate}
\end{lemma}

\begin{proof}  

Suppose that $(a,c)$ is implied by a component
$S$. \newline
\noindent \textit{First suppose $ac \notin E(G)$.} We show that $a$ and $c$ must have the same color. Suppose that this is not the case. Let $(a,d) \in S$ such that $(a,d) \rightarrow (a,c)$ or let $(b,c) \in S$ so that $(b,c) \rightarrow (a,c)$. In the former case $ad \notin E(G)$, $cd \in E(G)$, and hence $(a,d) \rightarrow (a,c) \rightarrow (a,d)$, which implies that $(a,c)$ is in $S$, a contradiction. On the other hand if $(b,c) \in S$ such that $(b,c) \rightarrow (a,c)$, we have $bc \notin E(G)$, and $ab \in E(G)$. Thus, we have $(a,c) \rightarrow (b,c) \rightarrow (a,c)$, and hence $(a,c) \in S$, a contradiction. Therefore, $a$ and $c$ have the same color. In this case, there is $(a,d) \in S$ so that $(a,d) \rightarrow (a,c)$, where $ad \notin E(G)$, and $cd \in E(G)$. 

Now we must have $N(a) \subseteq N(c)$ as otherwise, if $a$ has a neighbor $a'$ where $ca' \notin E(G)$, then $(a,c) \rightarrow (a',c) \rightarrow (a',c) \rightarrow (a',d) \rightarrow (a,d) \rightarrow (a,c)$, a contradiction. 

Since $(a,d) \in S$, there is some $(a,b) \rightarrow (a,d)$ in $S$ or
$ (b,d) \rightarrow (a,d)$. In the former case we have $cd,bd \in E(G)$, and $ad,ab \notin E(H)$, and hence, $(a,d) \rightarrow (a,b) \rightarrow (a,d)$ which proves that (1) occurs.  

If $(b,d) \in S$ so that $(b,d) \rightarrow (a,d)$ then $ad \in E(G)$. Let $(b,e) \in S$ such that $(b,e) \rightarrow (b,d)$. Observe that $be \not\in E(G)$ and $de \in E(G)$. Now it is easy to see that $(b,d) \rightarrow (a,d) \rightarrow (a,e) \rightarrow (b,e) \rightarrow (b,d)$. This shows that (2) occurs.  

\noindent \textit{Second suppose $ac \in E(G)$.} In this case we should have some $(a,d) \in S$ so that $(a,d) \rightarrow (a,c)$ with $dc \in E(G)$, and $ad \not\in E(G)$. Since $(a,d) \in S$, there is either $(a,d) \rightarrow (b,d) \in S$ or $(a,d) \rightarrow (a,b) \in S$. Suppose $(a,d) \rightarrow (b,d) \in S$. We have $bd \notin E(G)$, $ab \in E(G)$, and hence $(b,d) \rightarrow (a,d) \in S$. Furthermore, we should have  $bc \in E(G)$, otherwise, $(b,c) \rightarrow (b,d) \rightarrow (b,c)$, and $(a,c) \rightarrow (b,c)$, $(a,d) \rightarrow (a,c)$ implying that $(a,c) \in S$.  a contradiction. Therefore, we have a path $b,a,c,d$ with $N(a) \{c\} \subseteq N(c)$, and hence (3) is established. 

Now assume that $(a,d) \rightarrow (a,b) \in S$. We have $bd \in E(G)$, and $ad \notin E(G)$ and hence (4) occurs.  
\end{proof} 

We emphasize that $ab$ and $de$ from Lemma \ref{implyap} are independent edges.
Inclusion $N(a) \subseteq N(c)$ implies the following corollary.

\begin{cor}\label{source-sinkap}
If there is an arc from a component $S$ of $G^+$ to a pair $(x,y)
\notin S$ then $(x,y)$ forms a trivial component of $S$ that is a
sink component. If there is an arc to a component $S$ of $G^+$
from a pair $(x,y) \notin S$ then $(x,y)$ forms a trivial component of
$G^+$ that is a source. In particular, if there is a directed path in $G^+$ from component $S_1$ to component $S_2$, then $S_1=S_2$. 
\end{cor}

\section{Interval {\boldmath$k$}-graph recognition}

In this section, we present our algorithm for the recognition of interval
bigraphs. Firstly, to describe the algorithm, we introduce some technical definitions.

\begin{dfn} (envelope, $N^*[D]$)
Let $R$ be a set of pairs of $G^+$. The \emph{envelope} of $R$,
denoted $N^*[R]$, is the smallest set of pairs that contains $R$ and is closed under 
both reachability and transitivity (if $(u,v),(v,w) \in N^*[R]$ then $(u,w) \in N^*[R]$). 
\end{dfn}

\noindent{\textbf {Remark and {\boldmath$N_l^*[D]$} definition}} For the purposes of the proofs, we visualize taking the envelope of $R$
as divided into consecutive \emph{levels}, where in the zero-th level
we just replace $R$ by its reachability closure, and in each
subsequent level we replace $R$ by the rechability closure of its
transitive closure. 
The pairs in the envelope of $R$ can be
thought of as forming the arc of a digraph on $V(G)$, and each pair can be
thought of as having a label corresponding to its level. The pairs (arcs of the digraph) 
in $R$, and those implied by $R$ have label $0$, arcs obtained
by transitivity from the arcs labeled $0$, as well as all arcs
implied by them have label $1$, and so on. More precisely, $N^*[R]$ is the disjoint union of $R^0, 
R^1,\dots, R^k$, where  $R^0=N^+[R]$ (level zero), and each $R^i$ (level $i\ge 1$) consists of every
pair $(u,v)$ such that either $(u,v)$ is obtainable by transitivity in $R^{i-1}$ (meaning that there 
is some sequence $(u,u_1),(u_1,u_2),\dots,(u_{r-1},u_r),(u_r,v)$ in $R^{i-1}$), or $(u,v)$ is 
dominated by a pair $(u',v')$ obtainable by transitivity in $R^{i-1}.$ 
Let $N^*_l[D]= \bigcup_{i=0}^{i=l} R^i$. 
Note that $R \subseteq N^+[R] \subseteq N^*[R]$. 

\begin{dfn}[dictator component]
Let $\mathcal{R}=\{ R_1 , R_2,\dots,R_k, S\}$ be a set of
components of $G^+$ such that $N^*[\bigcup_{A\in \mathcal{R}} A]$ contains a
circuit. We say $S$ is a \emph{dictator} if for every subset $W$ of  $\mathcal{R} \setminus \{S\}$, there exist a circuit in the envelope of $(\bigcup_{A
\in W' } A)\cup (\bigcup_{B
\in \mathcal{R}\setminus W}  B)$, where $W'=\{ R'_i \mid R_i \in W\}$. In other words, $S$ is a dictator if by replacing some of the $R_i$s with $R'_i$s in $\mathcal{R}$ and taking 
the envelope of the union of elements we still get a circuit.
\end{dfn}

\begin{dfn}[complete set]
A set $D_1 \subseteq V(G^+)$ is called \emph{complete} if for every pair of coupled components 
$R,R'$ of $G^+$, exactly one of $R \subseteq D_1$ and $R' \subseteq D_1$ holds. 
\end{dfn}

A component $S$ is a dictator if and only if the envelope of every complete set $D_1$ containing $S$ has a circuit. 

\begin{algorithm}[p]
\caption{Algorithm for recognition of interval $k$-graphs}
  \label{alg-main}
  \begin{algorithmic}[1]
    \Function{Interval-k-graph}{$G$}
    \State {\textbf {Input:}} A connected $k$-partite graph $G$ 
    \State {\textbf {Output:}} An ordering  of $V(G)$ without patterns in Figure~\ref{fig:mix-patterns} or return 
    false
    
    \State Construct the pair-digraph $G^+$ of $G$, and compute its
components; 

if any component is  self-coupled  report false

\Statex \textbf{ \hspace{1mm} Stage1 : Adding (non-trivial strong) components}\label{stage1}
\State Initialize $D$ to be the empty set \label{line5}
    \State let $v_1,v_2,\dots,v_n$ be an ordering of the vertices of $G$ such that $i<j$ implies $c(v_i)<c (v_j)$

    \While { $\exists S_{v_i,v_j}$ and $S_{v_j,v_i}$, $i<j$ components in $G^+ \setminus D$}
    
     
     \If { $D \cup N^+[S_{v_i,v_j}]$ does not have a circuit } 
      \State add $N^+[S_{v_i,v_j}]$ to $D$, remove $N^+[S_{v_j,v_i}]$ from further consideration in this 
      step 
      
      \Comment { add $X$ to $D$  means add all the pairs of $X$ into $D$ }
      \ForAll { $(x,y) \in N^+[S_{v_i,v_j}]$ } $Dic(x,y)=S_{v_i,v_j}$
      \EndFor 
     \Else {}  
     \If { $D \cup N^+[S_{v_j,v_i}] $ does not have a circuit } 
       \State add $N^+[S_{v_j,v_i}]$ into $D$, delete $N^+[S_{v_i,v_j}]$ from further consideration  here 
       \label{line12}
       \ForAll {$(x,y) \in N^+[S_{v_j,v_i}]$ } set $Dic(x,y)=S_{v_j,v_i}$
      \EndFor
     \Else{} report that $G$ is not an interval $k$-graph
     \EndIf 
     
     \EndIf  
     
   \State \textbf{increase} $i$ by one

    \EndWhile

\Statex \textbf{ \hspace{1mm}  Stage2 : Computing  $N^*[D]$ and detecting dictator components}\label{stage2}

    \State Set $En=N^*[D]$, and $\DT=\varnothing$ \label{line15} \Comment{ $\DT$ is a set of components} 
    
    \While { $\exists (x,y) \in En \setminus D$ } \Comment{we consider the pairs in $En$ level by level}
      
      \State Move $(x,y)$ into $D$ and set $Dic(x,y)=$ \Call{Dictator}{$x,y,D$} 
      \If { $D \cup \{(x,y)\}$ contains a circuit } 
       add $Dic(x,y)$ into $\DT$ \label{line18}  
       
       \Comment{ $(x,y)$ is a complex pair }
      \EndIf
    \EndWhile

\Statex \textbf{ \hspace{1mm}  Stage3 : Adding dual of dictator components}\label{stage3}

      \State Let $D_1= \varnothing$ \label{line19}
      \ForAll{components $S \in  \DT$ }  
         add
$N^+[S']$ into $D_1$ \label{line20}
      \EndFor

   \ForAll { components  $R \in D \setminus \DT$ }
add $N^+[R]$ into $D_1$

\EndFor

      \State Set $D= N^*[D_1]$ \label{line22}
      \If {there is a circuit in $D$} 
      report $G$ is not an interval $k$-graph
      \EndIf
 \Statex \textbf{ \hspace{1mm}  Stage4 : Adding remaining trivial components, returning an ordering }\label{4}
     
     \While{ $\exists$ trivial component $S$ outside $D$, and $S$ is a sink component } 
      \State Add $S$ into $D$ and remove $S'$ from further consideration \label{line25}
 
    \EndWhile 

\Statex \textbf{ \hspace{1mm}  Output the final ordering }
     \ForAll { $(u,v) \in D$ } set $u \prec v$  
     
     \Comment{
      yielding an ordering of $V(G)$ without the
patterns from Figure 1}
     
     \EndFor

     \State Return the ordering $v_1 \prec v_2 < \dots \prec v_n$ of $V(G)$

    \EndFunction

\end{algorithmic}

\end{algorithm}

\begin{algorithm}

\begin{algorithmic}[H]

\Function{Dictator}{$x,y,D$}

 \If{$(x,y) \in N^+[S]$ for some component $S$ in $D$ } return $S$
 
 \EndIf 
 
 \If {$x,y$ have different colors and 
 $(u,y) \in D$  dominates $(x,y)$} 
 
   return \Call{Dictator}{$u,y,D$} \Comment{ we mean the earliest pair $(u,y)$} 
 \EndIf
 
 \If {$x,y$ have the same color and 
 $(x,w) \in D$ dominates $(x,y)$ } 
 
   return  \Call{Dictator}{$x,w,D$}
 
 \EndIf
 
 \If {$x,y$ have the same color and $(x,y)$ is by transitivity on
 
  $(x,w),(w,y) \in D$} return \Call{Dictator}{$w,y,D$}
 
 \EndIf
 
 \If {$x,y$ have different colors
and $(x,y)$ is by transitivity on

  $(x,w),(w,y) \in D$} return  \Call{Dictator}{$x,w,D$} \label{line6}
 \EndIf 
 
  \EndFunction 
    
  \end{algorithmic}

\end{algorithm}

\begin{dfn}[simple pair, complex pair]
A pair $(x,y) \in G^+$ is \emph{simple} if it belongs to $N^+[S]$ for
some component $S$, otherwise we call it \emph{complex}. 
\end{dfn}

\begin{dfn}\label{minimal-circuit}
Let $D$ be a complete set and let $C$ be a circuit in $N^*[D]$. We say $C$ is a \emph{minimal circuit } if first, the latest pair in $C$ is created as early as possible (the smallest possible level) 
 during the execution of $N^*[D]$; second, $C$ has the minimum length; third, no pair in $C$ is by transitivity. 
\end{dfn}

\noindent \textbf{Definition of color {\boldmath$c(u)$}}. 
For every vertex $u$ of $G$ we denote by $c(u)$ index of
the partite class containing $u$. This is the color of $u$; $c(u)=k$ if $u \in V_k$.

\paragraph{High level overview of the algorithm} 
The algorithm begins by constructing $ G^+$. If any component of $ G^+$ is self-coupled, then 
$ G$ is not an interval k-graph. Next, we initialize an empty set $D$ (which will store selected pairs). 
If we add a pair \( (x,y) \) into \( D \), it means that in the final ordering, \( x \) must appear before \( y \).

The core of the algorithm involves selecting components of \( G^+ \) based on the following principles: 

\begin{itemize}
    \item If \( (x,y) \in D \), then \( (y,x) \) must be discarded,
    \item If \( (x,y) \in D \) and \( (x,y)(x',y') \in A(G^+) \), then \( (x',y') \) must also be in \( D \), Thus, once a component \( (x,y) \) is added to \( D \), all pairs within its corresponding component \( S_{x,y} \) are also included in \( D \).

    \item If \( (x,y), (y,z) \in D \), then \( (x,z) \) must also be in \( D \).
\end{itemize}

\noindent{Stage 1: Selecting Components:}
We first compute an ordering \( v_1, v_2, \dots, v_n \) of the vertices of \( G \) such that if \( v_i < v_j \), then the color of \( v_i \) is the same as or smaller than the color of \( v_j \). The algorithm proceeds in steps \( i = 1,2, \dots \), where at each step, we consider components \( S_{v_i,v_j} \) (with \( i < j \)) for inclusion in \( D \). The selection follows these rules: 
If adding \( S_{v_i,v_j} \) to \( D \) (along with its outgoing neighbors \( N^+[S_{v_i,v_j}] \)) does not create a \emph{circuit}, then we include \( N^+[S_{v_i,v_j}] \) in \( D \) and discard \( S_{v_j,v_i} \), Otherwise, we try to add \( N^+[S_{v_j,v_i}] \) to \( D \). If this results in a \emph{circuit}, then \( G \) is not an interval \( k \)-graph. If neither \( S_{v_i,v_j} \) nor \( S_{v_j,v_i} \) can be added to \( D \), then \( G \)  contain an \emph{exobiclique} or the forbidden structures in Figure \ref{fig:so1}.\\

\noindent {Stages 2 and 3: Closure and Dictator Components:}
Next, we compute the \emph{closure} of \( D \), ensuring that for each pair of components \( S \) and \( S' \), exactly one of them is in \( D \). If a circuit \( C \) is found in \( N^*[D] \), we identify a dictator component \( S \), which cannot be included in any complete set. Consequently, we remove all these dictator components \( S_1, S_2, \dots, S_t \) from \( D \) and define a new set \( D_1 \), which includes their dual components \( S'_1, S'_2, \dots, S'_t \) along with all other elements of \( D \setminus (S_1 \cup S_2 \cup \dots \cup S_t) \). If \( N^*[D_1] \) contains a \emph{circuit}, then \( G \) is not an interval \( k \)-graph; otherwise, we update \( D \) as \( D_1 \) and set \( D = N^*[D] \).\\

\noindent{Stage 4: Handling Trivial Components}
Finally, we add any remaining trivial components from \( G^+ \setminus (D \cup D') \) (where \( D' \) is the \emph{dual} of \( D \)), starting with the \emph{sink components}. This step does not introduce \emph{circuits}, so we do not need to check for conflicts.
At the end of the algorithm, we derive a final order \( \prec \) setting \( u \prec v \) whenever \( (u,v) \in D \).

\subsection{Proof of the correctness of the Algorithm}

\noindent \textbf{Proof of Theorem \ref{th:main}}
The correctness of Algorithm 1 follows from Lemma \ref{lem:correctness}. We denote the degree of a vertex $z$ of $H$ by $d_z$.
In order to construct the digraph $G^+$, we need to list all
neighbors of each pair in $G^+$. If the vertices $x$ and $y$ in $G$ have different colors,
then the pair $(x,y)$ of $G^+$ has $d_y$ out-neighbors; and if $x$ and $y$ have the same color, then
the pair $(x,y)$ has $d_x$ out-neighbors in $G^+$. 
For a fixed vertex $x$ with $c(x)=0$, the number
of all pairs that are neighbors of all pairs $(x,z)$,  $z
\in V(H)$, is  $nd_x+d_{y_1}+d_{y_2}+\dots +d_{y_n}$, where
$y_1,y_2,...,y_n$ are all of different colors than $c(x)$. We
can use a linked list structure to represent $G^+$, therefore, overall, it takes time
$\mathcal{O}(mn)$ to construct $G^+$. Notice that in order to
check whether a component $S$ is self-coupled, it suffices to choose any pair $(a,b)$ in $S$ and check if $(b,a)$ is also in $S$. 
The latter task can be done in time $\mathcal{O}(mn)$, using Tarjan's strongly-connected component algorithm. Since we maintain a partial
order on $D$, once we add a new pair to $D$, we can decide whether that pair
closes a circuit or not. Computing $N^*[D]$ also takes time $\mathcal{O}(n(n+m))=\mathcal{O}(mn)$
since there are $\mathcal{O}(mn)$ edges in $G^+$ and 
$\mathcal{O}(n^2)$ vertices in $G^+$. Note that the envelope of $D$ is computed at most twice.

Once a pair $(x,y)$ is added to $D$, we put an arc from $x$ to $y$ in the partial order and give the arc $xy$ a time label (also called level). Once a circuit is formed at Stage 2,  we can find a dictator component $S$ using the \Call{Dictator}{} function, and store $S$ in the set $\mathcal{DT}$.
Therefore, we spend at most $\mathcal{O}(nm)$ time to find all the dictator components. Stage 4, in which we add the remaining pairs, takes time at most $\mathcal{O}(n^2)$. 
Therefore, the overall running time of the algorithm is $\mathcal{O}(nm)$.  

We start by giving some technical definition that are used in the correctness proof. 

\begin{dfn}
Let $(x,y)$ be a pair in $D$ by transitivity at the (earliest) level $l \ge 0$. By a \emph{minimal chain} between $x$ and $y$ we mean a sequence $(x_0,x_1),(x_1,x_2),\linebreak \dots,(x_{n-1},x_n)$ of minimum length ($n$) of pairs in $D$ with $x_0=x$ and $x_n=y$, such that each $(x_i,x_{i+1}) \in D$ for $0 \le i \le n-1$ at some level before $l$ by reachability (and not by transitivity). We also say $(x_0,x_n)$ is by transitivity on the minimal chain $(x_0,x_1),(x_1,x_2),\dots,(x_{n-1},x_n)$.
\end{dfn}

\begin{dfn}
Let $\CH=(x_0,x_1),\dots,(x_{n-1},x_n)$ be a minimal chain between $x_0$ and $x_n$ in $N_l^*[D]$, $0 \le l$. We say that pair $(x_i,x_{i+1})$ is a \textbf{tail} pair if there exists $(x_i,a_{i+1}) \in N_l^*[D]$ such that $(x_i,a_{i+1}) \rightarrow (x_i,x_{i+1})$ with $x_{i+1}a_{i+1} \in E(G)$, and $x_ia_{i+1} \notin E(G)$ ($x_i$, $a_{i+1}$ have different colors). On the other hand, we say $(x_i,x_{i+1})$ is a \textbf{head} pair if there exists $(a_i,x_{i+1}) \in N_l^*[D]$ so that $(a_i,x_{i+1}) \rightarrow (x_i,x_{i+1})$ with $a_ix_i \in E(G)$ and $x_ix_{i+1} \notin E(G)$, and $a_ix_i \in E(G)$, and $x_i,x_{i+1}$ have different colors.   
\end{dfn}

Now we are ready to prove the following correctness lemma for our algorithm. 

\begin{lemma}\label{lem:correctness}
Algorithm $1$, produces an ordering of the vertices of $G$ without the forbidden patterns depicted in Figure \ref{fig:forbidden-pattern} if and only if $G$ is an interval $k$-graph. 
\end{lemma}
\begin{proof}
Suppose we encounter a circuit while creating $N^*[D]$. The main ingredient of the proof is to assume this circuit is minimal. That is, $C$ contains a pair that added to $N^*[D]$ was the earliest pair, and 
no pair within the circuit arises purely by transitivity. Among possible choices, $C$ is assumed to have the shortest length. Since transitive closure is applied, we first analyze minimal chains of the form $CH: (x,y_1),(y_1,y_2),\dots,(y_{n-1},y_n),(y_n,y)$, 
where if each pair belongs to $N^*[D]$, then $(x,y)$ is also included in $N^*[D]$. 

Note that each pair in the circuit $C$, say $(x_i,x_{i+1})$, is derived from the reachability of some $(x,y)$, where $(x,y)$ originates by transitivity, and thus, it can be assumed that $(x,y)$ is 
formed via a minimal chain.
Lemma \ref{chain_1} in the appendix show some essential properties about a minimal CH, while a  sequence of Lemmas (\ref{head-tail},\ref{tail-tail},\ref{head-head},\ref{alternate-pairs}) examines consecutive 
pairs $(y_i,y_{i+1})$ and $(y_{i+1},y_{i+2})$ within a chain $CH$ and demonstrates that their type (head of tail) must alternate when the chain length is at least $4$.  
Furthermore, additional structural properties of minimal chains, 
proven in Lemmas \ref{chain_2}, \ref{chain_3}, \ref{chain_5}, and \ref{chain_4}, 
lead to the conclusion that the length of a minimal chain is at most 3 pairs. 
Consequently, the length of a minimal circuit is at most 4. In Theorem \ref{stage1-correctness-tm}, 
analyzing the presence of a circuit in Stage 1 of the algorithm reveals that such a circuit either 
detects an exobiclique (if its length is 4) or identifies a new obstruction (as shown in Figure \ref{fig:so1}) 
if its length is 3. This establishes the correctness of Algorithm 1 in Stage 1. 

Additional structural insights emerge if a circuit appears in Stage 2 of the algorithm. Lemma \ref{lem:dictator-component} establishes 
the existence of a component $X$ (so-called dictator component) such that, regardless of how other components are chosen in Stage 1, a circuit inevitably forms in Stage 2 as long as we keep $X$ in $D$. Thus, $X$ must not be included in $D$. If adding $X'$ and removing $X$ from $D$ also results in a circuit in $N^*[D]$, then $G$ cannot be an interval $k$-graph. 

\end{proof}

We present a series of lemmas discussing the structural properties of a minimal chain and minimal circuit during the computation of $N^*[D]$.

\begin{lemma}\label{chain_1}
Suppose a pair $(x,y) \in N_{l+1}^*[D]$ is obtained by a minimal chain $\CH= (x_0,x_1), (x_1,x_2),\linebreak[1] \dots, (x_{n-1},x_n), (x_n,x_{n+1})$ ($x_0=x$ and $x_{n+1}=y$) in $N_l^*[D]$, and there is no circuit formed in $N_{l+1}^*[D]$ by adding $(x,y)$. Suppose $x_i$ and $x_{i+1}$ have different colors and they are not adjacent. Then $x_{i+1}x_j \notin E(G)$.
\end{lemma}

\begin{proof}
Assume, for the sake of a contradiction, that $x_{i+1}x_j \in E(G)$. Since $j \neq i, i+1$, 
the arc $(x_i,x_{i+1})(x_i,x_j)$ exists in $G^+$. As $(x_i,x_{i+1}) \in N^*[D]$ 
and $(x_i,x_j) \in N^*[D]$, the chain 
$(x_{i-1},x_i), (x_i,x_j), (x_j,x_{j+1}), \dots,\linebreak[1] (x_{i-2},x_{i-1})$ when $j>i$ is a 
shorter chain, contradicting the minimality of $\CH$. Similarly, for $i<j$, circuit $(x_j,x_{j+1}), \dots,(x_{i-1},x_i),(x_i,x_j)$ is in $N^*[D]$, a contradiction to our assumption that the current $N^*[D]$ does not have a circuit.
 \end{proof}

\begin{lemma}\label{head-tail}
Suppose a pair $(x,y) \in N_{l+1}^*[D]$  is obtained by a minimal chain $\CH= (x_0,x_1),(x_1,x_2),\linebreak[1]\dots, (x_{n-1},x_n),\linebreak[1](x_n,x_{n+1})$ ($x_0=x$ and $x_{n+1}=y$) in $N_l^*[D]$, and there is no circuit formed in $N_{l+1}^*[D]$ by adding $(x,y)$. If $(x_i,x_{i+1})$ is a head pair and $(x_{i+1},x_{i+2})$ is a tail pair, then  $x_{i+1},x_{i+2}$ have the same color and different from the color of $x_i$. Furthermore, $x_ix_{i+2} \in E(G)$.
\end{lemma}

\begin{proof}

By definition, there exist $(a_i,x_{i+1}) \to (x_i,x_{i+1})$ in $N^*_l[D]$ 
with $a_ix_i \in E(G)$ and $x_ix_{i+1} \notin E(G)$. Similarly, 
there exist $(x_{i+1},a_{i+2}) \to (x_{i+1},x_{i+2})$ in $N^*_l[D]$ 
with $a_{i+2}x_{i+2} \in E(G)$ and $x_{i+1}x_{i+2} \notin E(G)$.
Now $a_ia_{i+2} \notin E(G)$, otherwise $(x_{i+1},a_{i+2}) \rightarrow (x_{i+1},a_i)$, a circuit in $N^*_l[D]$. 

For $l>1$, $x_ix_{i+2} \in E(G)$ must hold. Otherwise, independent edges $a_ix_i$ and $a_{i+2}x_{i+2}$ would imply that $S_{x_i,x_{i+2}}$ is a component, and should have been selected earlier according to the algorithm's rules, a contradiction to the minimality of $CH$. 

Now consider the case where $l=1$. Observe that $x_{i+1}x_{i+2} \notin E(G)$ by Lemma \ref{chain_1}. Consider the case where $a_ix_{i+1} \notin E(G)$ and $c(a_i) \neq c(x_{i+1})$. 
Since $(a_i,x_{i+1}),(x_i,x_{i+1})$ belong to the same 
component and $(x_i,x_{i+1}) \in N^*_l[D]$, we 
conclude that $c(x_i) < c(x_{i+1})$ or $c(a_i) < c(x_{i+1})$. 
Additionally, since $(a_i,x_{i+2}),\linebreak[1](x_i,x_{i+2}),(a_i,a_{i+2}),(x_i,a_{i+2})$ belong to the 
same component, $S_{x_i,x_{i+2}}$ should have been selected 
before component $S$ where $(x_{i+1},x_{i+2}) \in N^+[S]$, leading to a shorter chain.


In the case where $c(a_i)=c(x_{i+1})$, there is some $b_{i+1}$ so that $x_ia_i$ and $b_ix_{i+1}$ are independent edges of $G$, meaning that $a_ib_{i+1},x_ix_{i+1} \not\in E(G)$. Observe that $(x_i,b_{i+1}), (a_i,b_{i+1}), (x_i,x_{i+1}), \linebreak[1] (a_i,x_{i+1})$ are in a same component. Furthermore, $b_{i+1}x_{i+2} \not\in E(G)$, as otherwise, $(x_i,b_{i+1}) \rightarrow (a_i,b_{i+1}) \rightarrow (a_i,x_{i+2}) \rightarrow (x_i,x_{i+2})$, and hence, $(x_i,x_{i+2}) \in N^+[S_{a_i,x_{i+1}}]$, a shorter chain. Now $(x_{i+1},x_{i+2})$ and $(x_i,x_{i+2})$ are in  components. As argued earlier, $S_{x_i,x_{i+2}}$ should have been selected before $S_{x_{i+1},x_{i+2}}$, 
contradicting the  minimality of $CH$.
\end{proof}

\begin{lemma}\label{tail-tail}
Suppose a pair $(x,y) \in N_{l+1}^*[D]$ is obtained by a minimal chain $\CH = (x_0,x_1),(x_1,x_2), \linebreak[1]\dots, (x_{n-1},x_n),\linebreak[1](x_n,x_{n+1})$ ($x_0=x$ and $x_{n+1}=y$) in $N_l^*[D]$, and there is no circuit formed in $N_{l+1}^*[D]$ by adding $(x,y)$. If $(x_i,x_{i+1})$ is a tail pair and $(x_{i+1},x_{i+2})$ is a tail pair, then there exist $a_{i+1},a_{i+2}$ so that $a_{i+1}x_{i+1},a_{i+2}x_{i+2} \in E(G)$ and $ x_{i+2}a_{i+1},a_{i+2}x_i,a_{i+2}x_{i+1} \linebreak[1]  \notin E(G)$, and $S_{x_i,x_{i+1}}$ and $S_{x_i,x_{i+2}}$ are components. 

\end{lemma}

\begin{proof}
By definition, there exist $a_{i+1}x_{i+1},a_{i+2}x_{i+2}$ edges of $G $ such that $x_ia_{i+1},
\linebreak[1]x_{i+1}a_{i+2} \notin E(G)$. Now $a_{i+1}x_{i+2} \notin E(G)$, otherwise $(x_i,a_{i+1}) \rightarrow (x_i,x_{i+2})$, a shorter circuit. Notice that by minimality of $\CH$, $(x_i,x_{i+1})$ is not by transitivity. Now, let $(a_i,x_{i+1}) \in N^*_{l+1}[D]$ such that 
$(a_i,x_{i+1}) \to (x_i,x_{i+1})$. Since $a_ix_i \in E(G)$, 
we observe that $a_ix_{i+2} \notin E(G)$, as otherwise, 
the sequence $(a_i,a_{i+1}) \to (x_{i+2},a_{i+1}) \to (x_{i+2},x_{i+1})$ would 
form a shorter circuit. Since $a_ix_i$ and $a_{i+2}x_{i+2}$ are 
independent edges, $(x_i,x_{i+2})$ should have been chosen earlier, contradicting the minimality of $CH$.

Now, let $(a_i,x_{i+1}) \in N^*_{l+1}[D]$ with $(a_i,x_{i+1}) \rightarrow (x_i,x_{i+1})$. We have $a_ix_i \in E(G)$. Now it is easy to see that $a_ix_{i+2} \notin E(G)$ otherwise, $(a_i,a_{i+1}) \rightarrow (x_{i+2},a_{i+1}) \linebreak[1] \rightarrow (x_{i+2},x_{i+1})$, is a shorter circuit. 
Now $a_ix_i,a_{i+2}x_{i+2}$ are independent edges and hence $(x_i,x_{i+2})$ should have been chosen. 
Moreover, $a_ix_{i+1} \notin E(G)$ otherwise, $(x_{i+1},x_{i+2}) \rightarrow (a_i,x_{i+2}) \rightarrow (a_i,a_{i+2}) \rightarrow (x_i,a_{i+2}) \rightarrow (x_i,x_{i+2})$, a shorter circuit. These means we should have chosen $(x_i,x_{i+1})$ before $(x_{i+1},x_{i+2})$. 

If there is some $b_{i+1}$ so that $(x_i,b_{i+1}) \rightarrow (x_i,a_{i+1})$, then $S_{x_i,a_{i+1}}$ is in a component, and we can replace $x_{i+1}$ by $a_{i+1}$ in the chain $\CH$, meaning that the condition of the lemma holds for the replaced chain. 
\end{proof}

\begin{lemma}\label{head-head}
Suppose a pair $(x,y) \in N_{l+1}^*[D]$ is obtained by a minimal chain $\CH= (x_0,x_1),(x_1,x_2),\linebreak[1]\dots, (x_{n-1},x_n),\linebreak[1](x_n,x_{n+1})$ ($x_0=x$ and $x_{n+1}=y$) in $N_l^*[D]$, and there is no circuit formed in $N_{l+1}^*[D]$ by adding $(x,y)$. If $(x_i,x_{i+1})$ is a head pair and $(x_{i+1},x_{i+2})$ is a head pair, then  $x_{i},x_{i+2}$ have the same color,  $x_ix_{i+1},x_{i+1}x_{i+2} \notin E(G)$.  Furthermore, there exists $x_{i+1}a_{i+1} \in E(G)$ so that $a_{i+1}x_{i+2} \notin E(G)$, that is, $(x_{i+1},x_{i+2})$ is in a component. 
\end{lemma}
\begin{proof}
By definition, there exist $(a_i,x_{i+1}) \rightarrow (x_i,x_{i+1})$ in $N^*_l[D]$ with $a_ix_i \in E(G)$, and $x_ix_{i+1} \notin E(G)$. In addition, there exist $(a_{i+1},x_{i+2}) \rightarrow (x_{i+1},x_{i+2})$ with $a_{i+1}x_{i+1} \in E(G)$, and $x_{i+1}x_{i+2} \notin E(G)$. We observe that $a_ix_{i+2} \notin E(G)$ otherwise $(x_{i+1},x_{i+2}) \rightarrow (x_{i+1},a_i)$, a shorter circuit.    
Notice that $(x_i,x_{i+1}) \rightarrow (x_i,a_{i+1})$. So we may replace the pairs $(x_i,x_{i+1})$, 
$(x_{i+1},x_{i+2})$ by $(x_i,a_{i+1})$, $(a_{i+1},x_{i+2})$, obtained chain $\CH'$. Notice that $(a_{i+1},x_{i+2})$ is not by transitivity; else $\CH'$ contradicts the minimality of $\CH$. Therefore, $(a_{i+1},x_{i+2})$ is in a component and by Lemma \ref{implyap}, $(x_{i+1},x_{i+2})$ is in a component. 

The first case is when $a_{i+1}$ and $x_{i+2}$ have different colors and $a_{i+1}x_{i+2} \notin E(G)$. If $c(x_i) \ne c(x_{i+2})$, then by Lemma \ref{chain_1}, $x_ix_{i+2} \notin E(G)$, and hence $(x_i,x_{i+2}),(a_i,x_{i+2})$ are in a component. According to the rules of the algorithm, $(x_i,x_{i+1})$ must be in a component and therefore, according to the rules of the algorithm $(x_i,x_{i+2})$ should have been chosen before the component $S_{x_{i+1},x_{i+2}}$, hence a shorter circuit. Therefore, $c(x_i)=c(x_{i+2})$, and we are done here. 

So we may continue by assuming that there exists $x_{i+2}a_{i+2} \in E(G)$ such that $a_{i+1}a_{i+2} \notin E(G)$. Now again, $x_ia_{i+2} \notin E(G)$, otherwise we have $(x_i,x_{i+1}) \rightarrow (x_i,a_{i+1})$, and $(a_{i+1},a_{i+2}) \rightarrow (a_{i+1},x_i)$, a shorter circuit. Now $S_{x_i,x_{i+2}}$ is in a component. As we argued in the previous case, we must have $c(x_i)=c(x_{i+2})$. 
\end{proof}

\begin{lemma}\label{alternate-pairs}
Suppose $(x,y) \in N_{l+1}^*[D]$ is obtained by a minimal chain $\CH= (x_0,x_1), (x_1,x_2),\linebreak[1]\dots, (x_{n-1},x_n), (x_n,x_{n+1})$ ($x_0=x$ and $x_{n+1}=y$) in $N_l^*[D]$, and there is no circuit formed in $N_{l+1}^*[D]$ by adding $(x,y)$. For $i< n-2$, if $(x_i,x_{i+1})$ is a head pair, then $(x_{i+1},x_{i+2})$ is a tail pair, and if $(x_i,x_{i+1})$ is a tail pair then $(x_{i+1},x_{i+2})$ is a head pair. 
\end{lemma}
\begin{proof}     
     According to Lemma \ref{tail-tail} we can assume that there is a chain $\CH'$ by replacing $a_{i+1}$ by $x_{i+1}$ in $CH$.  Now $(x_i,a_i)$ and $(a_i,x_{i+2})$ are tail and head, respectively. Thus, we consider the case where $(x_i,x_{i+1})$ is a head pair and $(x_{i+1},x_{i+2})$ is also a head pair. First, suppose $(x_{i+2},x_{i+3})$ is a tail pair. By Lemma \ref{head-tail}, we have $x_{i+1}x_{i+3} \in E(G)$. However, $(x_i,x_{i+1}) \rightarrow (x_i,x_{i+3})$, a shorter chain. Thus, we may assume $(x_{i+2},x_{i+3})$ is also a head pair. According to Lemma \ref{head-head} there exists $x_{i+1}a_{i+1} \in E(G)$ such that $x_{i+1}x_{i+2},a_{i+1}x_{i+2}  \notin E(G)$ and $x_{i+2}$ have different colors than $x_{i+1},a_{i+1}$.  According to Lemma \ref{head-head} there exists $x_{i+2}a_{i+2} \in E(G)$ such that $x_{i+2}x_{i+3},a_{i+2}x_{i+3}  \notin E(G)$ and $x_{i+3}$ have different colors than $x_{i+2},a_{i+2}$. Notice that $x_ix_{i+1} \notin E(G)$. Now $x_{i+1}a_{i+2} \notin E(G)$, otherwise $(x_{i+1},x_{i+2}) \rightarrow (x_i,a_{i+2})$, and hence we consider the chain $\CH'=(x_0,x_1),\dots,(x_i,a_{i+2}),(a_{i+2},x_{i+3}),\dots, (x_{n-1},x_0)$, which is shorter than $\CH$, a contradiction. Now $(x_{i+1},a_{i+2}),(x_{i+1},x_{i+2}),(a_{i+1},x_{i+2})$ are in a component. 
     
     Note that there exists $(a_i,x_{i+1}) \rightarrow (x_i,x_{i+1})$ in $N^*_l[D]$.
     Observe that $a_{i+2}a_i \notin E(G)$, otherwise, $(x_{i+1},a_{i+2}) \rightarrow (x_{i+1},a_i)$ in $N^*_l[D]$, a circuit in $N^*_l[D]$. With the same line of reasoning $x_{i+2}a_i,x_ix_{i+2} \notin E(G)$. Therefore, now $(x_i,x_{i+2})$ is in a component and according to the rules of the algorithm, we should have selected $(x_i,x_{i+2})$, no later than $(x_{i+1},x_{i+2})$, a shorter chain.  
\end{proof}

\begin{lemma}\label{chain_2}
Suppose $(x,y) \in N_{l+1}^*[D]$ is obtained by a minimal chain $\CH= (x_0,x_1), (x_1,x_2),\linebreak[1] \dots, (x_{n-1},x_n),\linebreak[1](x_n,x_{n+1})$ ($x_0=x$ and $x_{n+1}=y$) in $N_l^*[D]$, and there is no circuit formed in $N_{l+1}^*[D]$ by adding $(x,y)$. Suppose $x_ix_{i+1} \notin E(G)$ and $x_i,x_{i+1}$ have different colors. Then one of the following occurs:

\begin{itemize}
\item $x_ix_{i+2} \in E(G)$ and $x_{i+1},x_{i+2}$ have the same color. 
\item $n=2$, $x_i,x_{i+2}$ have the same color, and both $(x_{i+1},x_{i+2}),(x_i,x_{i+1})$ are in components. 
\end{itemize}
\end{lemma}

\begin{proof}
\textit{Case 1.} Suppose $(x_i,x_{i+1})$ is a head pair. 
Then, there exists $(y_{i+1},x_{i+1}) \to (x_i,x_{i+1})$ in $N^*[D]$ with $x_iy_{i+1} \in E(G)$. 
By Lemma \ref{chain_1}, we know that $x_{i+1}x_{i+2} \notin E(G)$. Furthermore, if $y_{i+1}x_{i+2} \in E(G)$, 
then $(x_{i+1},x_{i+2}) \to (y_{i+1},x_{i+1})$ results in $(y_{i+1},x_{i+1}) \in N^*[D]$, 
contradicting our assumption.\\

\noindent\textit{Subcase 1.1.} If $(x_{i+1},x_{i+2})$ is a tail pair, then $(x_{i+1},y_{i+2}) \to (x_{i+1},x_{i+2})$ 
exists in $N_l^*[D]$, where $y_{i+2}x_{i+2} \in E(G)$ and $x_{i+1}y_{i+2} \notin E(G)$. If $y_{i+2}y_{i+1} \in E(G)$, then $(x_{i+1},y_{i+2}) \to (x_{i+1},y_{i+1})$, 
contradicting the assumption that $(x_{i+1},y_{i+1}) \notin N_l^*[D]$. If $x_{i+2}x_i \notin E(G)$ or 
if $x_{i+1}$ and $x_{i+2}$ have different colors, then $x_iy_{i+1}$ and $y_{i+2}x_{i+2}$ 
form independent edges, implying $(x_i,x_{i+2})$ is in a strong non-trivial component and belongs to $D$, 
yielding a shorter chain.\\

\noindent\textit{Subcase 1.2.} If $(x_{i+1},x_{i+2})$ is a head pair, then $(y_{i+2},x_{i+2}) \to (x_{i+1},x_{i+2})$ exists in $N_l^*[D]$, 
where $y_{i+2}x_{i+1} \in E(G)$ and $y_{i+2}x_{i+2} \notin E(G)$. By Lemma \ref{alternate-pairs}, 
we must have $i \geq n-2$, implying $x_{i+3}$ does not exist.

If $x_{i-1}$ exists, then by Lemma \ref{alternate-pairs}, $(x_{i-1},x_i)$ is a tail pair, 
implying $(x_{i-1},y_i) \to (x_{i-1},x_i)$ in $N_l^*[D]$. If $y_ix_{i+2} \in E(G)$, 
then $(x_{i-1},y_i) \to (x_{i-1},x_{i+2})$, forming a shorter chain. 
Similarly, if $y_iy_{i+2} \in E(G)$, then $(y_{i+2},x_{i+2}) \to (y_i,x_{i+2})$ results in a contradiction. 
Consequently, $x_iy_i$ and $y_{i+2}x_{i+1}$ are independent edges, leading to an alternative shorter chain.
Additionally, if $(y_{i-1},y_i) \to (x_{i-1},y_i)$ in $N_l^*[D]$ and $y_{i-1}x_{i-1} \in E(G)$, 
then $y_{i-1}y_{i+2} \notin E(G)$. Otherwise, $(y_{i-1},y_i) \to (y_{i+2},y_i) \to (y_{i+2},x_i)$ 
introduces a circuit in $N_l^*[D]$. Since $x_{i-1}y_{i-1}$ and $x_{i+1}y_{i+2}$ are 
independent edges, $(x_{i-1},x_{i+1}) \in N_l^*[D]$, creating a shorter chain. Thus, 
we conclude that $n=2$ and $x_i, x_{i+2}$ have the same color.\\

\noindent\textit{Case 2.} If $(x_i,x_{i+1})$ is a tail pair, then $(x_i,y_{i+1}) \to (x_i,x_{i+1})$ in $N_l^*[D]$, 
where $x_iy_{i+1} \notin E(G)$, $x_{i+1}y_{i+1} \in E(G)$, and $x_i, x_{i+1}, y_{i+1}$ have 
distinct colors. By Lemma \ref{alternate-pairs}, we must have $i \geq n-2$, meaning $x_{i+3}$ does not exist. \\

\noindent \textit{Subcase 2.1.} If $(x_{i+2},x_{i+3})$ is a tail pair, then by Lemma \ref{alternate-pairs} $(x_{i+2},y_{i+3}) \to (x_{i+2},x_{i+3})$ 
in $N^*[D]$, where $x_{i+3}y_{i+3} \in E(G)$, $x_{i+2}y_{i+3} \notin E(G)$, and $x_{i+2}, y_{i+3}$ have 
different colors. By Lemma \ref{chain_1}, $x_{i+1}x_{i+3} \notin E(G)$. 
If $y_{i+1}y_{i+3} \notin E(G)$, then $x_{i+1}y_{i+1}$ and $x_{i+3}y_{i+3}$ are independent edges, 
leading to $(x_{i+1},x_{i+3})$ being in a component, forming a shorter chain.

If $x_{i-1}$ exists, and $(y_{i-1},x_i) \to (x_{i-1},x_i)$ in $N_l^*[D]$, where $y_{i-1}x_{i-1} \in E(G)$ and $x_{i-1}x_i \notin E(G)$, 
then $y_ix_{i-1} \notin E(G)$ and $y_{i-1}x_{i+1} \notin E(G)$. This results in $y_{i-1}x_{i-1}$ 
and $y_ix_{i+1}$ being independent edges, meaning $(x_{i-1},x_{i+1}) \in N_l^*[D]$, forming 
a contradiction similar to Subcase 1.2.

\end{proof}

\begin{lemma}\label{chain_3}
Suppose $(x,y) \in N^*[D]$ is obtained by a minimal chain $\CH= (x_0,x_1),(x_1,x_2),\linebreak[1]\dots,(x_{n-1},x_n),\linebreak[1](x_n,x_{n+1})$ ($x_0=x$ and $x_{n+1}=y$) in $N^*[D]$ at level $l$, and there is no circuit formed in $N^*[D]$ by adding $(x,y)$ (level $0$ to level $l$ pairs). Suppose $x_ix_{i+1} \notin E(G)$ and $x_i,x_{i+1}$ have different colors. Then $x_ix_{i+2} \in E(G)$ and if $x_{i+3}$ exists then we have $x_{i+2}x_{i+3} \notin E(G)$. 
\end{lemma}

\begin{proof}
    By Lemma \ref{chain_2} when $x_{i+3}$ exists we have $x_ix_{i+2} \in E(G)$.  By Lemma \ref{chain_1} (1), $x_{i+1}x_{i+3} \notin E(G)$. For contradiction suppose $x_{i+2}x_{i+3} \in E(G)$.
First assume $x_i,x_{i+3}$ have different colors. Now $x_ix_{i+3} \notin E(G)$, otherwise, $(x_i,x_{i+1}) \rightarrow (x_{i+3},x_{i+1})$ in $N_l^*[D]$,  and hence, $(x_{i+3},x_{i+1}) \in N_l^*[D]$, implying a  circuit $(x_i,x_{i+1}),( x_{i+1},x_{i+2}),\linebreak[1] (x_{i+2},x_{i+3}), (x_{i+3},x_i)$ in $N_l^*[D]$. Now we have $(x_{i+2},x_{i+3}) \linebreak[1] \rightarrow (x_i,x_{i+3})$ in $N_l^*[D]$ (because $x_{i+2}x_i \in E(G)$, $x_ix_{i+3} \notin E(G)$), and we get a shorter chain by passing $x_{i+1},x_{i+2}$. This shows that $x_i,x_{i+3}$ must have the same color. Let $(x_{i+2},y_{i+2}) \rightarrow (x_{i+2},x_{i+3})$ in $N^*[D]$ so that $x_{i+2}y_{i+2} \notin E(G)$ and $y_{i+2}x_{i+3} \in E(G)$ (notice that the other option is not possible since $x_{i+2}x_{i+3} \in E(G)$. Now, $x_iy_{i+2} \notin E(G)$, as otherwise, $(x_{i+2},y_{i+2})(x_{i+2},x_i)$ is an arc of $G^+$, and hence $(x_i,x_{i+1}),(x_{i+1},x_{i+2}),(x_{i+2},x_i)$ is a circuit. 
On the other hand, $(x_{i+2},y_{i+2})(x_i,y_{i+2})$ and $(x_i,y_{i+2})(x_i,x_{i+3})$ are arcs of $G^+$, and hence, we get a shorter chain bypassing $x_{i+1},x_{i+2}$, a contradiction. Therefore, $x_{i+2}x_{i+3} \notin E(G)$.
\end{proof}

\begin{lemma}\label{chain_5}
Suppose $(x,y) \in N^*[D]$ is obtained by a minimal chain $\CH= (x_0,x_1),(x_1,x_2),\linebreak[1]\dots, (x_{n-1},x_n),\linebreak[1](x_n,x_{n+1})$ ($x_0=x$ and $x_{n+1}=y$) in $N_l^*[D]$ and there is no circuit formed in $N_{l+1}^*[D]$ by adding $(x,y)$. Then for every $x_i$, ($i<n-1$) of $\CH$, $x_ix_{i+1} \notin E(G)$. 
\end{lemma}

\begin{proof}
    For contradiction suppose $x_ix_{i+1} \in E(G)$ where $x_i$ and $x_{i+1}$ have different color. We consider two cases. 

\noindent \textit{Case 1}. $x_{i+1},x_{i+2}$ have the same color. Now $x_{i}x_{i+2} \in E(G)$, as otherwise, $(x_{i+1},x_{i+2})(x_i,x_{i+2})$ is an arc of $G^+$, and hence $(x_i,x_{i+2}) \in N^*[D]$, a shorter chain bypassing $x_{i+1}$. Let $(x_i,y_i) \in N_l^*[D]$ such that $(x_i,y_i)(x_i,x_{i+1})$ is an arc of $G^+$, $x_iy_i \notin E(G)$, and $y_ix_{i+1} \in E(G)$. Let $(y_{i+1},x_{i+1}) \in N^*[D]$ such that $(x_{i+1},y_{i+1})(x_{i+1},x_{i+2})$ is an arc of $G^+$ with $y_{i+1}x_{i+2} \in E(G)$, and $x_{i+1}y_{i+1} \notin E(G)$. 
Now, $y_ix_{i+2} \notin E(G)$, as otherwise, $(x_i,y_i)(x_i,x_{i+2})$, and hence, $(x_i,x_{i+2}) \in N^*[D]$, a shorter circuit. Now $y_ix_{i+1},y_{i+1}x_{i+2}$ are independent edges. Thus, we have $(x_i,y_i)(x_{i+2},y_i)$ and $(x_{i+2},y_i)(x_{i+2},x_{i+1})$ are arcs of $G^+$, implying that $(x_{i+2},x_{i+1}) \in N_l^*[D]$, a contradiction. \\

\noindent \textit{Case 2}. $x_{i+1},x_{i+2}$ have different colors. By Lemma \ref{chain_2} $x_{i+1}x_{i+2} \in E(G)$. 

By Case 1 and Lemma \ref{chain_2}, $x_{i+2}x_{i+3} \in E(G)$. Let $(x_{i+2},y_{i+2}) \in N^*[D]$ such that there is an arc $(x_{i+2},y_{i+2})(x_{i+2},x_{i+3})$ in $G^+$ with $y_{i+2}x_{i+3} \in E(G)$ and $x_{i+2}y_{2+1} \notin E(G)$ ($x_{i+2},y_{i+2}$ have different colors). Notice that $y_{i+2}x_{i+1} \notin E(G)$, as otherwise, $(x_{i+2},y_{i+2})(x_{i+2},x_{i+1})$ is an arc of $G^+$. and $(x_{i+2},x_{i+1}) \in N^*[D]$, a contradiction.
Moreover, $y_ix_{i+3} \notin E(G)$, otherwise, $(x_i,y_i)(x_i,x_{i+3})$ is an arcs of $G^+$, and hence, $(x_i,x_{i+3}) \in N^*[D]$, a shorter circuit. Now $(x_{i+2},y_{i+2})(x_{i+1},y_{i+2})$, $(x_{i+1},y_{i+2})(x_{i+1},x_{i+3}) $ are arcs of $G^+$, and hence $(x_i,x_{i+3}) \in N^*[D]$, a shorter circuit. 
\end{proof}

\begin{lemma}\label{chain_4}
Suppose $(x,y) \in N^*[D]$ is obtained by a minimal chain $\CH= (x_0,x_1),(x_1,x_2),\dots,\linebreak[1] (x_{n-1},x_n),\linebreak[1](x_n,x_{n+1})$ ($x_0=x$ and $x_{n+1}=y$) in $N_l^*[D]$, and there is no circuit formed in $N_{l+1}^*[D]$ by adding $(x,y)$ (level $0$ to level $l$ pairs). Suppose $x_ix_{i+1} \notin E(G)$ and $x_i,x_{i+1}$ have different colors. Then the following hold. 
\begin{itemize}
  \item [1.] $n=2$ and $(x_i,x_{i+1}),(x_{i+1},x_{i+2})$ are in components.

  \item [2.] If $n>2$, $x_{i+1},x_{i+2}$ have the same color and $x_ix_{i+2} \in E(G)$.
    \item [3.] If $x_{i+3}$ exists then $x_{i+2}x_{i+3} \notin E(G)$ and $x_i,x_{i+3}$ have the same color and $x_{i+1},x_{i+2}$ have the same color different from the color of $x_i$.
   \item [4.] $n=3$.    
\end{itemize}
\end{lemma}

\begin{proof}
    If $x_{i+3}$ and $x_{i-1}$ do not exist, then by Lemma \ref{chain_2} $n=2$ and $(x_i,x_{i+1}),\linebreak (x_{i+1},x_{i+2})$ are in components. By Lemma \ref{chain_2} and Lemma \ref{chain_3} we have $x_ix_{i+2} \in E(G)$. Suppose $x_{i+3}$ exists.
We show that $x_{i+2},x_{i+3}$ must have different colors. Suppose that this is not the case. Now we have $x_ix_{i+3} \in E(G)$, as otherwise, $(x_{i+2},x_{i+3}) \rightarrow  (x_i,x_{i+3})$ in $N_l^*[D]$, and hence, we get a shorter chain  bypassing $x_{i+1},x_{i+2}$. Let $(x_{i+1},y_{i+1})  \rightarrow  (x_{i+1},x_{i+2})$  in $N^*[D]$  with $y_{i+1}x_{i+2} \in E(G)$ and $x_{i+1}y_{i+1} \notin E(G)$. 
Now $y_{i+1}x_{i+3} \notin E(G)$, as otherwise, $(x_{i+1},y_{i+1}) \rightarrow (x_{i+1},x_{i+3})$ in $N_l^*[D]$, and hence, by passing $x_{i+2}$ we get a shorter chain. Now $(x_{i+2},x_{i+3}) \rightarrow (y_{i+1},x_{i+3})$ and $(y_{i+1},x_{i+3}) \rightarrow (y_{i+1},x_i)$ in $N_l^*[D]$, implying that $(y_{i+1},x_i) \in N^*[D]$, and getting a circuit $(x_i,x_{i+1}),(x_{i+1},y_{i+1}),(y_{i+1},x_i)$ in $N_l^*[D]$, a contradiction. 
Therefore, $x_{i+2},x_{i+3}$ must have different colors and $x_i, x_{i+3}$ have the same color. This proves 3. 

Suppose $x_{i+4} \ne x_i$ exists. By Lemma \ref{chain_1}, $x_{i+3}x_{i+4} \notin E(G)$ and $x_ix_{i+4} \notin E(G)$. 
First, assume that $x_{i+3},x_{i+4}$ have different colors. Let $(y_{i+4},x_{i+4}) \in N^*[D]$, so that the arc $(y_{i+4},x_{i+4}) (x_{i+3},x_{i+4})$ is in $G^+$ (here $y_{i+4}x_{i+3} \in E(G)$). Notice that $x_iy_{i+4} \notin E(G)$, as otherwise, $(y_{i+4},x_{i+4})(x_i,x_{i+4})$ would be a shorter chain bypassing $x_{i+1},x_{i+2},x_{i+3}$. Now, $x_ix_{i+2},y_{i+4}x_{i+3}$ are independent edges, and hence $(x_i,x_{i+3})$ is in a strong component. Therefore, it has been selected to be in $D$, yielding a shorter circuit. 

Second, let us assume $x_{i+3}$ and $x_{i+4}$ have the same color. Thus, there is $(x_{i+3},y_{i+4}) \in N^*[D]$, so that $(x_{i+3},y_{i+4})(x_{i+3},x_{i+4})$ is an arc of $G^+$. Observe that $y_{i+4}x_i \notin E(G)$, otherwise, $(x_{i+3},y_{i+4})(x_{i+3},x_i)$ is an arc of $G^+$,  yielding a  circuit $(x_i,x_{i+1}),(x_{i+1},x_{i+2}),(x_{i+2},x_{i+3}),\linebreak[1]
(x_{i+3},x_i)$. Now $x_{i+2}x_{i+4}$ must be an edge of $G$, as otherwise $x_ix_{i+2},x_{i+4}y_{i+4}$ are independent edges, and hence $(x_i,x_{i+4})$ is in a strong component and hence $(x_i,x_{i+4})$ must be in $D$, according to the rules of the algorithm. Therefore, $n=3$. 
\end{proof}

\begin{lemma}\label{number-of-tail-pairs}
Let $C=(x_0,x_1),(x_1,x_2),\dots,(x_{n-1},x_n),(x_n,x_0)$ be a minimal circuit in $N_{l}^*[D]$. 
Then the number of tail pairs in $C$ is two and $n \le 3$.  
\end{lemma}

\begin{proof}
First suppose more than two tail pairs in $C$. Let $(x_i,x_{i+1}),\linebreak[1](x_j,x_{j+1}),(x_k,x_{k+1})$, $0 \le i \le j \le k \le n$ (here $x_{n+1}=x_0$) be tail pairs in $C$. Notice that by Lemma \ref{alternate-pairs} $j-i>1$ and $k-j>1$.  Since $(x_i,x_{i+1}), (x_j,x_{j+1}), (x_k,x_{k+1})$ are tail pairs, there are $(x_i,a_{i+1}) \rightarrow (x_i,x_{i+1})$, $(x_j,a_{j+1}) \rightarrow (x_j,x_{j+1})$ and $(x_k,a_{k+1}) \rightarrow (x_k,x_{k+1})$ in $N^*_l[D]$ with $a_{i+1}x_{i+1},a_{j+1}x_{j+1},a_{k+1}x_{k+1} \in E(G)$, and $x_ia_{i+1},x_ja_{j+1},x_ka_{k+1} \notin E(G)$. Let $c(x)$ denote the partite set of $x$ (color of the vertex $x \in G$). Suppose $c(x_{i+1}) \ne c(a_{j+1})$ and $c(a_{i+1}) \ne c(x_{j+1})$. Now $a_{i+1}x_{j+1} \notin E(G)$ otherwise $(x_i,a_{i+1}) \rightarrow (x_i,x_{j+1})$ in $N^*_l[D]$ a shorter circuit. Similarly $a_{j+1}x_{i+1} \notin E(G)$. Now $(x_{i+1},x_{j+1})$ is in a component and, according to the rules of the algorithm, it should have been selected no later than $(x_{i+2},x_{i+3})$, and hence a shorter circuit in $N_l^*[D]$. 
Therefore, by a similar argument the following should occur. 

\begin{itemize}
    \item $c(a_{i+1})=c(x_{j+1})$ ( or $c(x_{i+1})=c(a_{j+1})$) 
    \item $c(a_{j+1})=c(x_{k+1})$ ( or $c(x_{k+1})=c(a_{j+1})$)
    \item 
    $c(a_{i+1})=c(x_{k+1})$ ( or $c(x_{i+1})=c(a_{k+1})$ )
\end{itemize}
So we may assume $c(a_{i+1})=c(x_{j+1})$ ( or $c(x_{i+1})=c(a_{j+1})$). First, suppose $c(a_{j+1})=c(x_{k+1})$. Since $x_{j+1}a_{j+1}$ is an edge, $c(x_{j+1})=c(a_{i+1}) \ne c(x_{k+1})$. Thus, we must have $c(a_{k+1}=c(x_{i+1})$. Note that $a_{k+1}a_{j+1} \notin E(G)$, otherwise $(x_k,a_{k+1}) \rightarrow (x_k,x_{j+1})$, a shorter circuit in $N_l^*[D]$. Now $x_{j+1}x_{k+1}$ is an edge of $G$, otherwise $(x_{j+1},a_{k+1}),(x_{j+1},x_{k+1})$ are in a component, and hence according to the rules of the algorithm, $(x_{j+1},x_{k+1}) \in N_l^*[D]$, a shorter circuit. By Lemma \ref{alternate-pairs}, $(x_{j+1},x_{j+2})$ is a head pair, and hence $x_{j+1}x_{j+2} \notin E(G)$. By Lemma \ref{chain_1}, $x_{j+2}x_{k+1} \notin E(G)$. Now, $(x_{j+1},x_{j+2}),(x_{k+1},x_{j+2})$ are in a same component, a shorter circuit (note that $j+2 \ne k+1$). By Lemma \ref{alternate-pairs} we conclude that $n \le 3$.
\end{proof}

\begin{theorem}\label{stage1-correctness-tm}
If in stage 1 of the algorithm we encounter a component $S$
such that we cannot add $N^+[S]$ and $N^+[S']$ to the current $D$,
then $G$ has an exobiclique as an induced subgraph, and
it is not an interval $k$-graph.
\end{theorem}

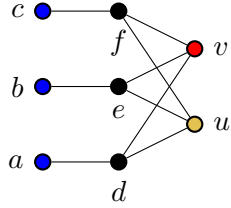
\begin{figure}[hbtp]
  \centering
  \begin{tikzpicture}[scale=1.0]
    \node[a, label=right:$u$] (u) at (3,1.5) {}; 
    \node[r, label=right:$v$] (v) at (3,2.5) {};
    \foreach[count=\y] \m/\n in {a/d, b/e, c/f} {
      \node[b, label=left:$\m$] (\m) at (1,\y) {}; 
      \node[s, label=below:$\n$] (\n) at (2,\y) {}; 
      \draw (\m)--(\n)--(u)  (\n)--(v);
    }
 \end{tikzpicture}
  \caption{Smaller obstruction}
  \label{fig:so2}
\end{figure}

\begin{proof}
The inability to add $N^+[S]$ and $N^+[S']$ arises because their 
inclusion creates circuits in $D \cup N^+[S]$ and $D \cup N^+[S']$, 
respectively. Suppose that adding $N^+[S]$ with $S = S_{v_i,v_j}$ to $D$ 
results in a minimal circuit $C=(x_0,x_1), (x_1,x_2), \dots , (x_{n-1},x_n), (x_n,x_0)$ in $D$.

By Lemma \ref{number-of-tail-pairs}, we have $n \le 3$. First, consider the case where $n=3$. Assume that $(x_0,x_1)$ and $(x_2,x_3)$ are head pairs and $(x_1,x_2)$ and $(x_3,x_0)$ are tail pairs. By Lemma \ref{head-tail} we have $x_0x_2 \in E(G)$, and $c(x_1)=c(x_2)$ and $x_0x_1 \notin E(G)$. Furthermore, $c(x_3)=c(x_0)$ have the same color and $x_2x_3 \notin E(G)$ ($c(x_2) \ne c(x_3)$). Note that there is $a_0$ such that $a_0x_0 \in E(G)$, $x_0x_1 \notin E(G)$. There is also $x_2a_2 \in E(G)$ so that $x_1a_2 \notin E(G)$. Note that $x_0a_2, a_0a_2 \notin E(G)$.  

Notice that $c(a_2) \ge  \min \{c(x_0),c(a_0)\}$ otherwise, since $a_0a_2,x_0a_0$ are not edges of $G$, $S_{a_2x_0}$ is in a component and we should have selected $(a_2,x_0)$ before $(x_0,x_1)$, implying a shorter circuit. We show that $c(x_0)=c(a_2)$ or $c(a_0)=c(a_2)$. Otherwise, $(x_0,a_2)$ is in a component, and therefore $(x_0,a_2)$ has been selected. Now $(x_0,a_2) \rightarrow (x_0,x_2)$, and hence we get a shorter circuit. \\

\textbf{Case 1:} Suppose $a_2b_2 \in E(G)$ where $x_1b_2 \notin E(G)$ and $a_2b_2 \in E(G)$. 
This implies that $a_0b_2 \notin E(G)$. Now, since $x_0a_0$ and $a_2b_2$ are 
independent edges, $(x_0, a_2)$ is part of a component. We also observe that $c(a_2), c(b_2) > c(x_1)$; 
otherwise, $(a_2, x_1)$ should have been chosen earlier. \\

\textbf{Subcase 1.1:} $x_1a_0 \notin E(G)$ and $c(a_0) \ne c(x_1)$. Now $(x_0,x_1)$ is in a component and hence $c(x_0)<c(x_1)$. Now $(x_0,a_2)$ must have been selected before $(x_1,a_2)$ and hence $(x_0,a_2) \rightarrow (x_0,a_2) \rightarrow (x_0,x_2)$ a shorter circuit.\\

\textbf{Subcase 1.2:}  $x_1a_0 \notin E(G)$, and $c(a_0)=c(x_1)$. Now $(a_0,x_1)$ is in a component and hence, there is $x_1b_1$ so that $a_0b_1 \notin E(G)$. Note that we must have $c(a_2)=c(x_0)$ and $c(b_2)=c(x_0)$ but this is a contradiction, as it would imply that we should have chosen $(a_2,x_1)$ because $c(x_1) > c(a_2)$ or $c(x_1) > c(b_2)$. \\

\textbf{Case 2.} There are $a_2b_2,x_1a_1 \in E(G)$ so that $x_1a_2,a_1b_2 \notin E(G)$. Note that $(x_0,x_1) \rightarrow (x_0,a_1)$. Now, $x_0b_2 \notin E(G)$, otherwise, $(a_1,b_2) \rightarrow (a_1,x_0)$ a shorter circuit. 

Note that again using the arguments in Case 1, we conclude that there are $x_1b_1 \in E(G)$ so that $a_0b_1 \notin E(G)$. Since $x_0x_2$ is an edge, $x_2b_1$ is  an edge of $G$, otherwise $(x_0,b_1) \rightarrow (x_2,b_1) \rightarrow (x_2,x_1)$.

Analogously, for pairs $(x_2,x_3),(x_3,x_0)$ we conclude that there are independent edges $x_2c_2,x_3a_3,uv$ of $G$ where $x_2x_3, c_2a_3, x_2u,a_3u,c_2v,x_3v \notin E(G)$. Furthermore, we have $vx_0 \in E(G)$. Now it is easy to see that $va_1,va_2,a_3a_1,a_3a_2$ are edges of $G$, otherwise we get a shorter circuit. However, an exobiclique appears on $G[\{x_0,x_1,x_2,x_3,a_0,a_1,c_2,b_2,a_3,u,v\}]$.

For $n < 3$, we conclude $n = 2$, where $(x_0, x_1)$ belongs to a component, 
and $(x_1, x_2)$ is implied by a component. Independent edges $x_0a_0, x_1a_1 \in E(G)$ 
exist such that $x_0x_1, a_0a_1 \notin E(G)$ and $c(x_0) \neq c(x_1)$. 
Moreover, edges $x_1b_1, c_2b_2 \in E(G)$ exist such that $x_1c_2, b_1b_2 \notin E(G)$ 
and $x_2b_1, x_2c_2, x_2a_1, x_2x_0 \in E(G)$. Suppose $(x_2, u) \rightarrow (x_2, x_0)$ in $D$. 
Then, $x_2, u_2 \notin E(G)$ and $x_0u_2 \in E(G)$. Here, $c(x_2) \neq c(u_2)$. 
Using similar arguments, we show that $G$ contains a forbidden structure as 
illustrated in Figure \ref{fig:so2}, confirming that $G$ is not an interval $k$-graph.

Now suppose $(x_2,u) \rightarrow (x_2,x_0)$ in $D$. Thus, we should have $x_2,u_2 \not\in E(G)$, and $x_0u_2 \in E(G)$ (assuming that $(x_2,x_0)$ is a tail pair). Here $c(x_2) \ne c(u_2)$. Notice that $a_1u_2 \in E(G)$, otherwise, $(x_0,a_1) \rightarrow (u_2,a_1) \rightarrow (u_2,x_2)$, a contradiction as $(x_2,u_2) \in D$. By similar argument, we conclude that $c_2u_2 \in E(G)$. Now $(x_0,x_1) \rightarrow (x_0,b_1) \rightarrow (u_2,b_1) \rightarrow (u_2,x_2)$ if $b_1u_2 \not\in E(G)$. Thus, $x_0u_2,a_1u_2,b_1u_2,c_2u_2 \in E(G)$. However, $G[\{x_0,x_1,x_2,a_0,a_1,b_1,b_2,c_2,u_2\}]$ is isomorphic to Figure 5 which is an obstruction to interval $k$-graphs. 
\end{proof}

\begin{lemma}\label{second-stage-circuit}
    Let $C=(x_0,x_1),(x_1,x_2),\dots,(x_{n-1},x_n),(x_n,x_0)$  be a minimal circuit formed in $N^*[D]$ in Stage 2 of the algorithm. Then $n=3$, and $x_0,x_3$ have the same color and $x_1,x_2$ have the same color and different from the color of $x_0$. Furthermore, $x_0x_2 \in E(G)$.
\end{lemma}

\begin{proof}
    Suppose $(x_n,x_0)$ is the last pair added to $N^*[D]$. Now $(x_0,x_1),(x_1,x_2),\dots, \linebreak[1] (x_{n-1},x_n)$ is a minimal chain between $x_0,x_n$. Thus, by Lemma \ref{chain_4}, either $n=2$ and $(x_0,x_1),(x_1,x_2),(x_2,x_0)$ are all components, resulting in a circuit in Stage 1 of the algorithm, or we have $n=3$, and $x_1,x_2$ have the same color and different from the color of $x_0$. Furthermore, $x_0x_2 \in E(G)$. 
\end{proof}

From Lemma \ref{chain_4} we derived the following Lemma. 

\begin{lemma}\label{chain1ap}
Let $(x,y)$ be a pair in $D$ after Stage 1 of the algorithm, and assume that the current $D$ has no circuit.

\begin{itemize}
\item Suppose $x$ and $y$ have the same color and $(x,w)(x,y) \in A(G^+)$ such that $(x,w)$ is by transitivity with a minimal chain $(x,w_1),(w_1,w_2),\dots,(w_m,w)$.
Then $m=2$ and vertices $x$ and $w_1$ have the same color and opposite to the color of $w_2$ and $w$.

\item Suppose $x$ and $y$ have different colors and $(w,y)(x,y) \in A(G^+)$ such that $(w,y)$ is in a 
trivial component. Then $(w,y)$ is implied by transitivity from a minimal chain
$(w,w_1), (w_1,w_2), (w_{2},y)$ where $w_1$ and $w_2$ have the same color,
opposite to the color of $w$ and $y$.
\end{itemize}
\end{lemma}

\begin{lemma}\label{lem:dictator-component}
Let $C: (x_0,x_1),(x_1,x_2),(x_2,x_3),(x_3,x_0)$ be a circuit formed at Stage 2 of Algorithm \ref{alg-main}. Then there is a component $S \in D$, so for any complete set $D_1$ containing $N^*[D_1]$ contains a circuit. 
\end{lemma}
\begin{proof}
By Lemma \ref{second-stage-circuit}, assume that $x_0,x_3$ has the same color and is opposite to the color of $x_1,x_2$. 
We first show that if one of the pairs $(x_1,x_2)$ is not in $N^+[S]$ for any component $S$, then it is implied by $(x_1,w)$ where $(x_1,w)$ is implied transitivity $(x_1,w_1),(w_1,w_2),(w_2,w)$ where $x_1,w_1$ have the same color and are opposite to the color of $w_2,x_0$. The same happens for the other pairs of the circuit. In other worlds, the pairs involved in creating such a circuit are a subset of $V_i \times V_j$ for some fixed $1 \le i < j \le k$. Suppose $(x_1,w) \in N_l^*[D] \rightarrow (x_1,x_2)$ with $wx_2 \in E(G)$, and $x_1w \notin E(G)$. We observe that $wx_3 \notin E(G)$, else $(x_1,w) \rightarrow (x_1,x_3)$, contradicting the minimality of the circuit. This implies that if $w,x_3$ has different colors, then $(x_2,x_3) \rightarrow (w,x_3) \rightarrow (x_2,x_3)$ in $N_1^*[D]$. Now we get a circuit $(x_0,x_1),(x_1,w),(w,x_3),(x_3,x_0)$, which contradicts the minimality of the circuit and Lemma \ref{second-stage-circuit}. Therefore, $w,x_0,x_3$ have the same color. Now, by Lemma \ref{chain1ap}, $x_1,w_1$ have the same color and different from the color of $w_2,x_3$. Using the same argument, we can show that the vertices involved in creating circuit $C$ belong to $V_i \cup V_{i+1}$ (two different color classes). Therefore, we can apply Lemmas 6.14, 6.15, 6.16, 6.17, 6.18, and 6.19 from \cite{arash-jgt} the lemma is proved. 
\end{proof}

\section{Generalization}

For a graph $H$ and a coloring $c\colon V(G) \to V(H)$, the graph $G$ equipped with 
intervals $I_v$ for each $v \in V(G)$ is an \emph{interval $H$-graph} if, for different 
$u,v \in V(G)$, there is an edge $uv \in E(G)$ if and only if $I_u \cap I_v \ne \emptyset$ 
and $c(u)c(v) \in E(H)$. Since interval $k$-graphs are interval $K_k$-graphs,
the concept of interval $H$-graphs generalizes interval $k$-graphs. 
Let $\chi(G)$ and $\omega(G)$ denote the chromatic number of $G$ and the maximum size
of a clique in $G$. For all intervals $H$-graphs we have $\chi(G) \le \chi(H)$ and
$\omega(G) \le \omega(H)$. Moreover, every graph $H$ is an interval $H$-graph where all intervals $I_v$ coincide.
In order to decide if $G$ is an interval $H$-graph we construct the auxiliary digraph $G^+$, as follows. The vertex set of $G^+$ consists of ordered pairs $(u,v)$ where $u \neq v \in V(G)$. The arc set of $G^+$ is defined as follows:
\begin{itemize}
    \item $(u,v)(u,v')$ is an arc if $uv \notin E(G)$, $c(u) \neq c(v)$, $c(v)c(v') \in E(H)$, and $vv' \in E(G)$.
    \item $(u,v)(u',v)$ is an arc if $c(u)c(u') \in E(H)$, $uu' \in E(G)$, and $vu' \notin E(G)$.
\end{itemize}

Clearly, if a component of $G^+$ contains a circuit, then $G$ is not an interval $H$-graph.  Assuming that no strong component of $G^+$ contains a circuit, the key challenge is to select, 
from each pair of components $S$ and $S'$, one component to be included in the set $D$, ensuring that $D$ remains closed under reachability and transitivity. We believe that our approach for interval $k$-graphs could be extended to this setting. However, we may end up having a new set of obstructions.

\end{document}